\documentclass[conference]{IEEEtran}
\IEEEoverridecommandlockouts

\usepackage{lipsum}
\usepackage{listings}
\usepackage{tikz}
\usepackage{multirow}
\usepackage{tikz}
\usepackage{cite}
\usepackage{booktabs}
\usepackage{hyperref}
\usepackage{cleveref}

\makeatletter
\let\old@lstKV@SwitchCases\lstKV@SwitchCases
\def\lstKV@SwitchCases#1#2#3{}
\makeatother
\usepackage{lstlinebgrd}
\makeatletter
\let\lstKV@SwitchCases\old@lstKV@SwitchCases

\lst@Key{numbers}{none}{%
    \def\lst@PlaceNumber{\lst@linebgrd}%
    \lstKV@SwitchCases{#1}%
    {none:\\%
     left:\def\lst@PlaceNumber{\llap{\normalfont
                \lst@numberstyle{\thelstnumber}\kern\lst@numbersep}\lst@linebgrd}\\%
     right:\def\lst@PlaceNumber{\rlap{\normalfont
                \kern\linewidth \kern\lst@numbersep
                \lst@numberstyle{\thelstnumber}}\lst@linebgrd}%
    }{\PackageError{Listings}{Numbers #1 unknown}\@ehc}}
\makeatother

\newcommand{\tool}{\textsc{cu\-Thermo}}
\newcommand{\smem}{\textsc{SMEM}}
\newcommand{\gmem}{\textsc{GMEM}}
\newcommand{\reg}{\textsc{REG}}
\newcommand{\squarechar}[1]{\tikz[baseline=(X.base)] \node[draw, minimum size=1em, inner sep=0pt] (X) {#1};}
\newcommand{\circlednum}[1]{\tikz[baseline=(char.base)]{
            \node[shape=circle, draw, inner sep=0.5pt] (char) {#1};}}

\begin{document}

\title{\tool{}: Understanding GPU Memory Inefficiencies with Heat Map Profiling}
\author{
\IEEEauthorblockN{
Yanbo Zhao\IEEEauthorrefmark{1}, Jinku Cui\IEEEauthorrefmark{1},
Zecheng Li\IEEEauthorrefmark{1}, Shuyin Jiao\IEEEauthorrefmark{1},
Xu Liu\IEEEauthorrefmark{1}, Jiajia Li\IEEEauthorrefmark{1}
}
% \IEEEauthorblockA{\IEEEauthorrefmark{1}North Carolina State University \\\{yzhao62, jcui23, zli94, sjiao2, xuliu88, jiajia.li\}@ncsu.edu}
\IEEEauthorblockA{\IEEEauthorrefmark{1}North Carolina State University \\
\url{yzhao62@ncsu.edu}, \url{jcui23@ncsu.edu}, \url{zli94@ncsu.edu}, 
\url{sjiao2@ncsu.edu}, \url{xuliu88@ncsu.edu}, \url{jiajia.li@ncsu.edu}}
}

% \subtitle{Category B: Experimental paper}
% CUDA Memory Thermometer
% Understanding Memory Inefficiencies with Memory Heat Map

\maketitle

\begin{abstract}

GPUs have become indispensable in high-performance computing, machine learning, and many other domains. Efficiently utilizing the memory subsystem on GPUs is critical for maximizing computing power through massive parallelism.  
Analyzing memory access patterns has proven to be an effective method for understanding memory bottlenecks in applications. However, comprehensive runtime and fine-grained memory profiling support is lacking on GPU architectures.  
In this work, we introduce \tool{}, a lightweight and practical profiling tool for GPU memory analysis. It operates on GPU binaries without requiring any modifications to hardware, operating system, or application source code.  
Given a CUDA application, \tool{} identifies memory inefficiencies at runtime via a heat map based on distinct visited warp counts to represent word-sector-level data sharing and provides optimization guidance in performance tuning iterations. Through our experiments on six applications, we identified five memory access patterns that are portable across different GPU architectures. 
% evaluating the portability of insights we find and the corresponding optimization strategies on two different GPUs, showing similar performance improvement range from $1.85\%$ to $721.79\%$ on both GPUs.
By evaluating optimization on two GPUs, \tool{} achieves up to $721.79\%$ performance improvement.

\end{abstract}

% \begin{CCSXML}
% <ccs2012>
%    <concept>
%        <concept_id>10010520.10010521.10010528.10010534</concept_id>
%        <concept_desc>Computer systems organization~Single instruction, multiple data</concept_desc>
%        <concept_significance>500</concept_significance>
%        </concept>
%    <concept>
%        <concept_id>10011007.10011006.10011066.10011069</concept_id>
%        <concept_desc>Software and its engineering~Integrated and visual development environments</concept_desc>
%        <concept_significance>300</concept_significance>
%        </concept>
%  </ccs2012>
% \end{CCSXML}

% \ccsdesc[500]{Computer systems organization~Single instruction, multiple data}
% \ccsdesc[300]{Software and its engineering~Integrated and visual development environments}

% \keywords{Profiling, GPU, memory, heat map, tool}

% \received{20 February 2007}
% \received[revised]{12 March 2009}
% \received[accepted]{5 June 2009}

\section{Introduction}\label{Section: introduction}

% \jli{The motivation part is weak. I feel DrGPUM's introduction section is not as strong as I expect. We can try to write a stronger one. The ``Intra-object inefficiency: oerallocation'' is very weak as the intra-object inefficiency I feel in DrGPUM work. }
% \jli{Could we have a table to compare this work with all previous work? I feel this work can stand out easily compared to others. If so, a table will be very attractive.}

As GPUs become more and more powerful, an increasing number of applications leverage their parallel processing capabilities to accelerate computations across various domains, including scientific computing, artificial intelligence, and data analytics. With the exponential growth of data, memory- and data-intensive applications have become a major component of GPU workloads, necessitating efficient memory management and optimization strategies.  Despite the growing accessibility of GPU programming frameworks such as CUDA~\cite{cuda} and OpenCL~\cite{opencl}, effectively utilizing GPUs to achieve high application performance remains a significant challenge for developers. These challenges stem from the complex programming model, intricate memory hierarchy, and the difficulty of efficiently mapping computations to hardware resources.

Effectively utilizing low-latency memory resources, such as registers, shared memory, and caches, is important for an application's performance. In particular, GPUs provide a configurable shared memory and L1 cache structure that shares the same physical space. This flexibility allows workloads to dynamically adjust cache behavior but also introduces additional complexity in performance tuning. Furthermore, the replacement policy of caches in modern GPUs is non-trivial and architecture-dependent~\cite{jia2018dissecting}, making it more difficult for programmers to predict cache behavior. The disparity in developers' familiarity with hardware intricacies further exacerbates these challenges, creating a gap between low-level memory optimization techniques and high-level application development.
Understanding memory access patterns during runtime, especially for input-sensitive applications, is crucial, to discover inefficiencies not covered by static analysis.
% as optimizing data locality and memory transactions can significantly enhance performance, reduce latency, and improve overall system efficiency.
% Despite the growing accessibility of GPU programming, effectively utilizing shared memory remains a challenging task for developers. Unlike traditional cache hierarchies in CPUs, 

To address memory optimization concerns, several GPU memory profiling tools have been developed. Vendor-provided profilers such as NVIDIA Nsight Compute~\cite{NsightCompute} and Compute Sanitizer~\cite{ComputeSanitizer} offer insights into memory bandwidth utilization, cache hit/miss rates, and memory allocation inefficiencies. However, these tools primarily provide high-level statistics, lacking fine-grained insights or reasons into the behavior of individual memory accesses. Additionally, academic profiling tools have explored various aspects of memory efficiency. GVProf~\cite{zhou2020gvprof} detects redundant memory accesses, while CUDAAdvisor~\cite{shen2018cudaadvisor} employs LLVM~\cite{lattner2004llvm} instrumentation to analyze memory reuse distance and divergence. While valuable, these tools are either limited to static analysis, dependent on compiler-based instrumentation, or lack runtime correlation between memory accesses and performance bottlenecks.

DrGPUM~\cite{lin2023drgpum} is a recent runtime profiler that introduces an object-centric approach to memory profiling, analyzing memory inefficiencies at both the macroscopic (object-level) and microscopic (intra-object) scales. It correlates memory inefficiencies with specific data objects and GPU API calls, providing actionable insights for developers. However, DrGPUM primarily focuses on identifying \textit{memory wastage}, such as unused or overallocated memory, rather than cache inefficiencies. It lacks explicit support for visualizing fine-grained memory access patterns within the streaming multiprocessor (SM) cores, a critical aspect in diagnosing performance issues related to shared memory and L1 cache interactions.
% \jli{Profiler does not give reasons, give behaviors. DrGPUM gives weak reasons.}

Compiler-based optimization frameworks attempt to improve memory efficiency by restructuring memory accesses, yet they are inherently constrained by memory safety models and conservative optimization strategies. Unlike human programmers, who can make domain-specific adjustments, compilers often lack the runtime information necessary to make aggressive transformations that exploit GPU memory hierarchy effectively. For example, memory access coalescing and shared memory utilization are highly workload-dependent and may require heuristics or application-specific tuning that a static compiler cannot fully anticipate. % \jli{Compilers or programmers do not well-utilize the GPU features. Our tool is a bridge from hardware memory to optimization oppotunities. better than static, drawback: limited by memory safety model so compiler cannot do aggressive optimization than users propose.}

The limitations of existing profilers and optimization frameworks highlight the need for real-time, hardware-aware memory profiling to assist both developers and compilers in understanding memory inefficiencies and making informed optimization decisions.  
However, to achieve this goal, three challenges need to be addressed.

\emph{Challenge 1:}~\emph{The traditional memory access counting method to represent hotness cannot identify inefficient memory access patterns.}
% \yanbo{3. Why don't use access account. 4. why warp\_id can measure the sharing level.}
The memory access counting method uses the number of memory accesses for addresses or cache lines to indicate how frequently a specific memory region is accessed.  
However, frequency information alone cannot distinguish data-sharing behavior among threads or warps on GPUs.  
For example, it fails to differentiate whether a memory region is accessed frequently by a single thread or shared and accessed by multiple threads within the same warp or across multiple warps.  
As a result, detecting inefficient memory access patterns becomes challenging.
\textbf{Since memory instructions are issued at the warp level, the number of distinct warps accessing a memory region serves as a reliable indicator of locality within a thread block.}

% Another significant advantage of \tool{} is its capability for \textit{runtime detection} of memory inefficiencies, ensuring that profiling accurately reflects input-sensitive program behavior. Static analysis techniques often struggle to account for the effects of varying inputs, leading to optimizations that may be overly conservative or ineffective in practice. By dynamically monitoring memory behavior during execution, \tool{} identifies inefficiencies that manifest under specific conditions, making optimization recommendations more reliable across diverse workloads.

\emph{Challenge 2:}~\emph{The granularity of memory regions should accurately reflect real memory access behavior on the hardware.}
There are different granularities of memory regions to consider, such as word, sector, cache line, and data object levels.  
Data objects are too coarse-grained to capture diverse behaviors across different code regions or CUDA kernels.  
At the word-level granularity, it is challenging to represent data-sharing behavior among threads and capture coalescing information accurately to reflect actual memory transactions.  
In NVIDIA’s cache architecture, a cache line comprises 128 bytes and is divided into four 32-byte \textit{sectors}~\cite{KernelProfilingGuide} for cache management.  
When a memory instruction is issued, the coalescing process consolidates targeted memory regions into sector-aligned requests.
\textbf{Sectors serve as the fundamental memory transaction units after hardware memory request coalescing, making sector-level memory access analysis critical for accurately diagnosing memory inefficiencies.}
% \textbf{Sectors serve as the fundamental memory transaction units after hardware memory request coalescing.  
% Analyzing memory access at the sector level is critical for accurately diagnosing memory inefficiencies. }
% \textbf{Sectors, as fundamental memory transaction units after coalescing, are critical for accurately diagnosing memory inefficiencies.}

\emph{Challenge 3:}~\emph{To collect the aforementioned information, nonselective trace collection generates massive and noisy traces, making accurate in-SM performance analysis impossible.}
An algorithm typically generates a large number of memory instructions, especially on GPUs, which execute massive numbers of threads, leading to an overwhelming trace volume.  
Additionally, traces from warps in different thread blocks share the same warp indexing information, polluting the trace data and making it difficult to accurately analyze program behavior.  
The sampling method must be carefully designed, as time-based sampling alone cannot effectively filter out noisy traces.  
\textbf{We introduce a thread block-sampling instrumentation strategy that records traces from only a single user specified thread block (default 0) as a representative, ensuring low runtime overhead while improving memory behavior modeling accuracy. And kernel sampling is also supported by a whitelist method to reduce overhead.}

We introduce \tool{}, a lightweight, runtime profiler that provides fine-grained insights into memory access behavior on NVIDIA GPUs, which records accessed warp\_ids within a target threadblock in the word-sector level, to monitor in-SM memory behavior. 
\tool{} implemented using NVIDIA’s NVBit \cite{villa2019nvbit} dynamic binary instrumentation framework, operates directly on compiled CUDA binaries, eliminating the need for source code modifications, recompilation, or additional compiler passes. This design ensures seamless integration into existing workflows, making it practical for both production environments and proprietary software.

In summary, this paper makes the following contributions:
\begin{itemize}
    \item We propose a novel runtime metric for categorizing memory behavior in GPU applications, by introducing the concept of the memory heat map. (\Cref{Sec:Methodology})
    \item We present a lightweight, modular tool that selectively samples memory instructions of a thread block to balance profiling accuracy and runtime overhead while providing extensibility to other GPU platforms. (\Cref{Sec:Implementation})
    % We present a \textit{lightweight, modular tool} that selectively samples the memory behavior of a thread block to balance profiling accuracy and runtime overhead. And provides extendibility to other GPU platforms. 
    \item We address five inefficient memory access patterns: hot spots, abuse of shared memory, false sharing, memory misalignment, and strided memory access through the memory heat map, enabling developers to pinpoint optimization opportunities. (\Cref{sec:patterns})
    % \item We address \textit{six inefficient memory access pattern: hot, random hot, overuse of shared memory, false sharing, memory misalignment and strided memory access in memory heat map}, enabling developers to pinpoint optimization opportunities.
    \item We evaluate inefficiency insights and apply optimizations to four applications on two different GPUs, achieving $1.85\%$ to $721.79\%$ performance improvement. (\Cref{Sec:Evaluation,sec:cases})
    % \item We evaluate inefficiency insights and make optimization in four applications in two different GPUs, achieving up to $721.79\%$ speedup in kernel cycles.
\end{itemize}

% \jli{Remove this paragraph later.}
% The remainder of this paper is organized as follows. \Cref{sec:background} provides background information on GPU memory hierarchy and profiling tools. \Cref{sec:design} details the design and implementation of \tool{}. \Cref{sec:evaluation} presents an experimental evaluation of our approach, and \Cref{sec:discussion} discusses potential limitations. Finally, \Cref{sec:conclusion} concludes the paper.

\section{NVIDIA GPU Background}\label{section:background}

% \jli{shorten}

% The numbered components correspond to key memory and execution units: (1) Registers, (2) Shared Memory, (3) L1 Cache. 

% \jli{Why DrGPUM does not need a background section?? I feel we need a background section, but we do NOT generally describe common knowledge of GPUs! Need to reduce this section. Need Yanbo's thoughts on this.} 

Modern NVIDIA GPUs are designed for massively parallel workloads, making them essential in fields such as scientific computing, artificial intelligence, and real-time graphics rendering\cite{CUDAProgrammingGuide}. 
This section gives a brief overview of the thread hierarchy of its programming model, memory hierarchy, and the relationship between them.
% \todo{change $\rightarrow$} and the popular NVBit binary instrumentation tool used in our proposed \tool{} workflow to help understand this work.

\subsection{Thread Hierarchy}
\begin{figure}[t]
    \centering
    \includegraphics[width=0.8\linewidth]{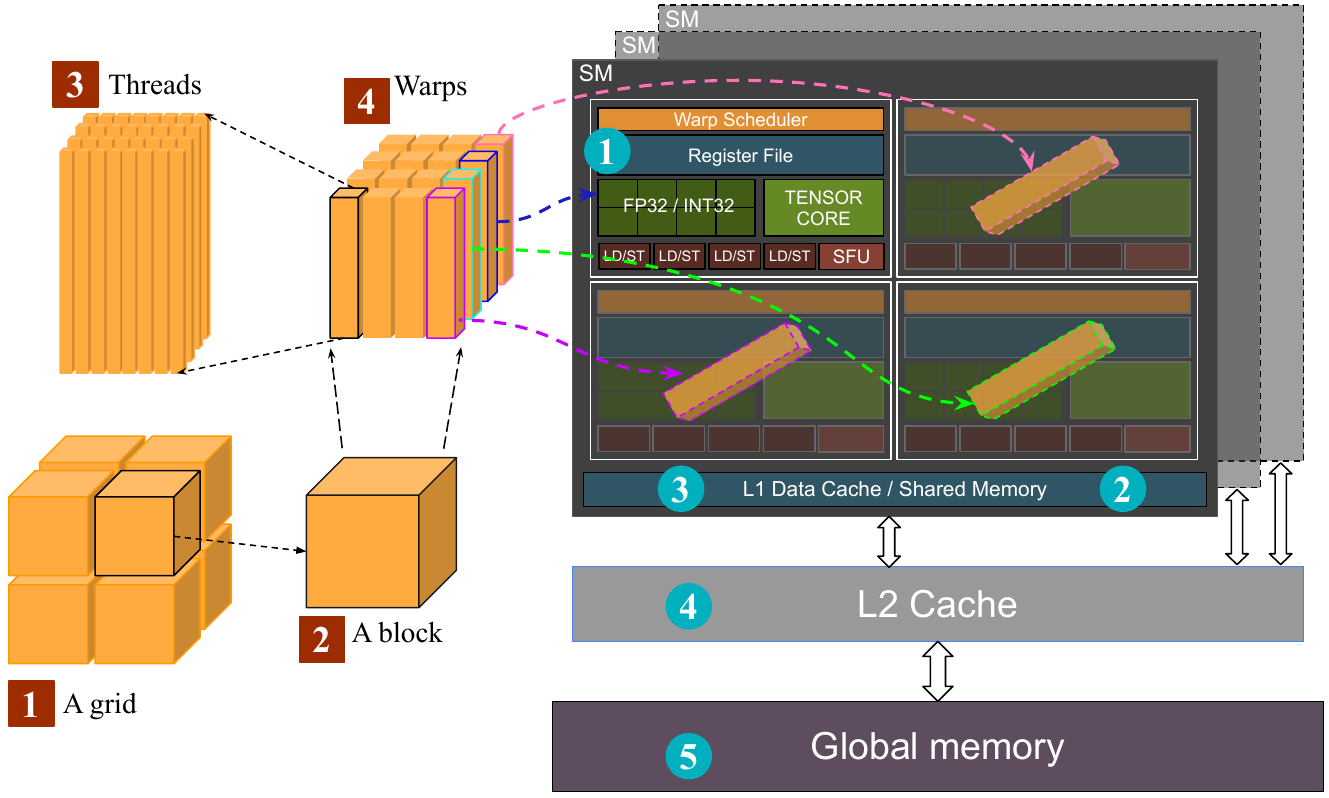}
    \vspace{-6pt}
    \caption{An abstract architecture of NVIDIA GPUs.}
    \label{fig:background}
    \vspace{-12pt}
\end{figure}

% Understanding the threading model is crucial for optimizing GPU performance. 
From a programmer's view, a single kernel in a CUDA program launches a grid \squarechar{1} that contains multiple thread blocks \squarechar{2}, each thread block consists of multiple threads \squarechar{3}. For an application to well utilize GPU's massive parallelism, usually tens of thousands or more threads will be allocated~\cite{CUDAProgrammingGuide}. 
From an architectural view, thread execution is organized into units called ``warps'' \squarechar{4}, each containing 32 threads that execute in a Single Instruction Multiple Threads (SIMT) fashion.
The core of the execution model is the Streaming Multiprocessor (SM), depicted in \Cref{fig:background}, which contains four sub-cores, including warp scheduler, execution units (CUDA cores, Tensor cores~\cite{VoltaWhitePaper}  and Special Function Units (SFU)), Load/Store units (LD/ST) and register files, and other components.
The warp scheduler issues instructions to warps that are ready for execution. While thread blocks are assigned to SMs in a non-deterministic manner with their mapping abstracted from users, the scheduling may vary across different runs. However, each thread block is always assigned to a single SM within a given execution. In this work, we analyze in-block memory access behavior as a reliable indicator for predicting the performance of in-SM memory components.

% The warp scheduler issues instructions to warps that are ready to execute; while thread blocks are assigned to SMs with their mapping hidden from users and non-deterministic, the assigned schedule can be totally different between different runs, but one thread block will only be attached to one SM in one execution. 
% In this work, we use in-block memory access behavior as an accurate perspective for predicting the performance of in-SM memory components.
% So, in-block memory access behavior is the most accurate perspective to predict in-SM memory components performance.
% and handled by the GPU driver.
% This variability can affect performance optimizations that rely on specific memory access patterns, making it important for developers to design memory-efficient algorithms that do not assume fixed scheduling.

\subsection{Memory Hierarchy}
The memory hierarchy commonly considered by CUDA programmers consists of registers (\reg{}) \circlednum{1}, shared memory (\smem{}) \circlednum{2}, L1 \circlednum{3} and L2 \circlednum{4} caches \footnote{We refer to data caches in this paper.}, and global memory (\gmem{}) \circlednum{5}.
The global memory is the slowest but largest memory component on a GPU device with latency around 1000 cycles if TLB miss or 350 cycles if TLB hit~\cite{jia2018dissecting}. L2 cache sits between SMs and the global memory, reducing latency to around 200 cycles and memory bandwidth pressure by caching frequently accessed data~\cite{jia2018dissecting}.
The memory hierarchy residing in an SM includes registers, shared memory and L1 cache.
Registers provide the fastest access (1 cycle) and are private to each thread, and GPUs have a large amount of registers (such as 256KB per SM for Ampere GPUs).
Shared memory and L1 cache share a configurable total size per SM (e.g., 128KB on \texttt{sm\_86}). Though they both have low latency (around 25 cycles~\cite{jia2018dissecting}) and are accessible to all threads within a thread block, shared memory is user-managed, while L1 cache is hardware-managed with around 20 cycles~\cite{jia2018dissecting}. Each memory block is aligned to 128 bytes and divided into four 32-byte sectors~\cite{KernelProfilingGuide} during data transfer between levels.
% Shared memory and L1 cache share a configurable total Static Random-Access Memory (SRAM) size per SM (e.g., 128KB on \texttt{sm\_86}). Though they both have low latency (around 25 cycles~\cite{jia2018dissecting}) and are accessible to all threads within a thread block, shared memory is user-managed, while L1 cache is hardware-managed with around 20 cycles~\cite{jia2018dissecting}. Each memory block is aligned to 128 bytes and divided into four 32-byte sectors~\cite{KernelProfilingGuide} during data transfer between levels.
% ; while the units to access shared memory is by \textit{bank} that is 32-byte wide.
The word size of NVIDIA GPUs is 32 bits (4 bytes). 

GPUs use massive warps to effectively hide memory latency by overlapping computation with memory operations. 
Thus, efficient CUDA program execution depends on aligning memory access patterns well with the hardware’s capabilities. 
% Moreover, memory access behavior is critical to program performance. 
Coalesced global memory access and efficient utilization of shared memory contribute to improved application performance~\cite{BestPractice}.
% bank conflict-free shared memory access 
% For instance, memory coalescing—where consecutive threads access contiguous memory addresses—can significantly enhance throughput by minimizing memory transaction and improving cache utilization.
Comprehending the intricate balance between thread execution, memory access, and scheduling behavior is challenging. Even advanced developers find it difficult to write an efficient CUDA program that fully utilizes GPU resources, including computational power and high memory bandwidth. Moreover, NVIDIA GPU architecture continues to evolve and improve with each generation. But time evaluated micro-architectures persists. 
We deploy the \textit{Word-Sector} granularity in this work to better abstract hardware behavior.
\section{Workflow Overview}\label{Sec:Workflow}

\begin{figure}[t]
    \centering
    \includegraphics[width=0.45\textwidth]{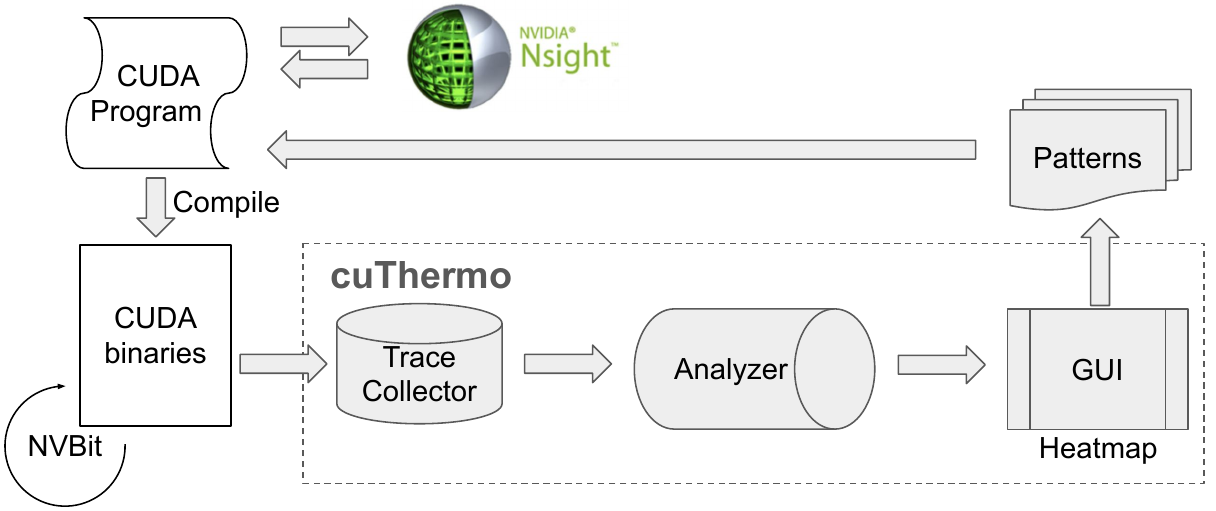}
    \vspace{-6pt}
    \caption{The performance tuning workflow leveraging \tool{}.}
    \label{fig:tool_workflow}
    \vspace{-2em}
\end{figure}

\tool{} is designed to be a lightweight and practical profiling tool for GPU memory analysis, illustrated in Figure~\ref{fig:tool_workflow}. 
It operates on GPU binaries without requiring any modifications to the hardware, operating system, or source code. 
Given a user's CUDA code, \tool{} identifies its memory access inefficiencies and recognizes access patterns during runtime. 
The patterns will be used by users to optimize the corresponding region of their CUDA program.
NVIDIA Nsight profiling tool can be used to testify the program behavior from a resource utilization perspective. 
Then the user could use \tool{} to detect the optimized version of the program again and check for other memory access inefficiencies, as the optimizing process iterates.

\tool{} consists of three major components: real-time trace collector, analyzer, and heat map visualization.
% It enables developers to optimize CUDA applications from .cu code by identifying memory access inefficiencies with minimal runtime overhead.
% , \tool{} provides a streamlined and efficient workflow for GPU memory profiling.
Trace collector employs NVBit-based dynamic instrumentation to collect fine-grained memory access data along with metadata such as program counters, access sizes, and active thread masks in a CUDA application. To mitigate profiling overhead and noise, a thread block-sampling approach is used to ensure that memory access patterns remain representative while minimizing performance interference. 
Analyzer processes the memory access data collected at runtime to generate key statistics about GPU memory usage. It extracts word-level memory access information and computes the number of distinct warps that have accessed each word and sector, revealing patterns of memory locality and contention. 
% By structuring the collected information into meaningful profiles, the analyzer transforms raw memory access logs into actionable insights.
The memory access logs are passed to the GUI visualization, where the memory footprints are aggregated and visualized as a heat map.
% , highlighting cache access inefficiencies. 
% The visualization tool provides valuable and intuitive insights to developers to pinpoint inefficiencies in memory behavior and provide guidance on memory pattern recognization and optimization suggestions.
The heat map highlights highly-shared memory regions, sparsely utilized sectors, and contention hot spots, providing developers with an intuitive way to diagnose performance bottlenecks.

% These methodological choices enable \tool{} to provide valuable insights into memory behavior while ensuring minimal
% The analyzer also identifies sparsely accessed memory regions and high-contention areas, enabling developers to pinpoint inefficiencies in memory allocation. 
% By leveraging binary instrumentation, \tool{} collects runtime memory access data and presents it in an intuitive heat map for developers to analyze memory efficiency. Figure~\ref{fig:tool_workflow} shows the workflow of \tool{}, which consists of a data collector, an analyzer, a profiler, and a GUI-based visualization system.

% \jli{Add portability.}
% \jli{These insights from \tool{}, combined with tailored solutions in \Cref{sec:patterns}, provide developers with a systematic approach to do GPU memory optimization and portable from different GPU architectures with similar SM design.
% Through \tool{}'s visualization, developers can identify and implement targeted optimizations. The effectiveness of these strategies is directly observable through subsequent heat maps. This approach bridges the gap between raw performance metrics and optimization strategies, allowing more algorithm developers to better understand the hardware architecture to optimize their algorithms.}

% , combined with tailored solutions in \Cref{sec:patterns},
These insights from \tool{} provide developers with a systematic approach to GPU memory optimization that remains portable across different GPU architectures. 
% \jli{Shorten the following sentences $\triangleright$}
As fundamental memory hierarchy principles persist across GPU generations, hardware-aware optimizations from \tool{}'s heatmaps are transferable between GPU families. 

\section{Methodology \& Implementation}

% \textcolor{green}{ku: data collector: how and why we sample 2. Implementation: how we wrap the target functions: cudaMalloc, cudaMallocHost, cudaMallocManaged, etc. 3. Patterns Definition 4. GUI design }

In this section, we introduce our heat map design and its construction techniques in \tool{}, then present five patterns that cannot be detected by the state-of-the-art tools.
% , then describe the techniques to construct the heat map. Finally, we present five patterns that cannot be detected using the state-of-the-art tools, such as NCU and DrGPUM. 
% traditional memory access counting method but can be identified through our heat map.

\subsection{Methodology}\label{Sec:Methodology}

Recognizing different access patterns has a significant impact on performance optimization in GPUs. For example, frequently accessed data used by a single thread could be stored in registers (\reg{}), which provide the fastest access (refer to \Cref{section:background}). Unlike CPUs, NVIDIA GPUs have a much larger register resource, making it possible to store high-frequency, low-sharing, thread-private data efficiently. 
Without recognizing this pattern, those memory regions will unnecessarily occupy precious L1 cache space, which has a longer access latency than \reg{} (refer to \Cref{section:background}), and will affect other data that genuinely requires L1 cache.

To detect inefficiencies in GPU memory, we define two essential components of a metric that are critical for revealing memory behavior: one captures the viewpoint that highlights inefficiencies, and the other defines the granularity that determines how the metrics are collected and organized.

% To detect inefficiencies in GPU memory, we introduce two orthogonal dimensions that are critical for revealing memory behavior: 
% To detect inefficiencies in GPU memory, two components of a metric that are critical for revealing memory behavior:
% one represents the perspective \jli{change the word} to reflect inefficiencies, and the other represents granularity which determines how the collected metrics are organized. 
% deciding which aspect to monitor

Previous works \cite{yoon2015memory,lin2023drgpum} use \textit{memory access counts} to analyze memory behavior: the metric is the number of memory accesses, while the granularity is per-word memory address or per data-object, indicating how frequently a specific memory region is accessed.
However, frequency information alone cannot distinguish data-sharing behavior among threads. 
For example, it fails to distinguish whether a memory region is accessed frequently by a single thread or simultaneously shared and accessed by multiple threads in one warp or even multiple warps. 
Other work \cite{koo2017access} enhanced memory access count by classifying intra-warp and inter-warp instructions to detect data sharing between warps, but their approach fails to classify the level of sharing between threads. This approach can only determine whether a memory region is shared, not how it is shared.

\textit{Example.} \Cref{fig:false_sharing} gives two different memory access patterns: perfect coalescing and false sharing~\cite{Will1993} (explained in \Cref{sec:patterns}). 
In \Cref{fig:false_sharing}(a), threads in a single warp access the same memory region, \texttt{sector 00}, once per thread. Its number of memory accesses is 1 for each word, 8 for the sector.
However, the memory access count in \Cref{fig:false_sharing}(b) is exactly the same with \Cref{fig:false_sharing}(a) but in a different memory access pattern. 
% shows multiple warps accessing the same sector with the same \textsc{memory access count} of 8, 
% , but no other warps share this region, indicating perfect coalescing. 
Simply counting the number of memory accesses does not differentiate between these two access patterns.
% reflect how a memory region being shared among threads
% from different warps. 
% This highlights that relying solely on access frequency misrepresents the sharing level. 

To address the issue of access count measurement, we propose the number of distinct warps as the \textsc{metric}, which describes data sharing among warps, and two levels of memory \textsc{granularity}---words and sectors---to reflect actual data transformation. We refer to our metric-granularity table as a heat map to represent the ``temperature'' of fine-grained memory regions, thereby revealing elusive inefficiencies arising from algorithm design and the mismatch between software and GPU hardware.

% \jli{Why using warps? 2. the memory transactions are submitted in warp-level, not thread level. 3. Not thread blocks, because the block binding is not-deterministic among multile runs. 3. \smem{}, L1 cache are shared among warps in one thread block, the threads within a warp could communicate via register shuffling. Thus, recognizing distinct warps information could reveal the parallel and data sharing patterns, that will be very helpful for performance optimization.}

% \subsubsection{Identifying Data Sharing Level}
\subsubsection{Metric---Distinct Warp Count}

GPU programming involves multiple levels of data sharing: among threads within the same warp, among threads from different warps within the same thread block, and among threads from different thread blocks. 
Different levels of sharing have varying memory access costs generally aligning with its corresponding memory hierarchies, aforementioned in \Cref{section:background}. 
% Threads within the same warp can use \textit{shfl} instructions for low-latency data exchange, typically completing in 4–10 cycles. Threads within a thread block but from different warps can leverage \smem{}, which usually incurs a latency of 20–30 cycles, though this often requires costly synchronization to prevent data races. Communication between threads in different thread blocks relies on \gmem{}, resulting in the highest overhead---hundreds of cycles.
Moreover, block binding varies across runs, making inter-block sharing unpredictable, except when data is extensively shared across all thread blocks. 
\emph{Thus, in this work, we focus on the first two levels of data sharing within the same thread block, which limits memory optimization to resources available within an SM.}
% Effectively leveraging the fastest memory resources plays a crucial role in optimizing overall application performance.
% These distinct memory demands, differing in both cost and memory hierarchy, play a crucial role in guiding optimization decisions.
% To identify different sharing levels in GPU programming, we first recall key concepts:   
% Sharing between threads can be categorized by cost and mechanism. 

\begin{figure}[h]
    \vspace{-0.9em}
    \centering
    \includegraphics[width=0.85\linewidth]{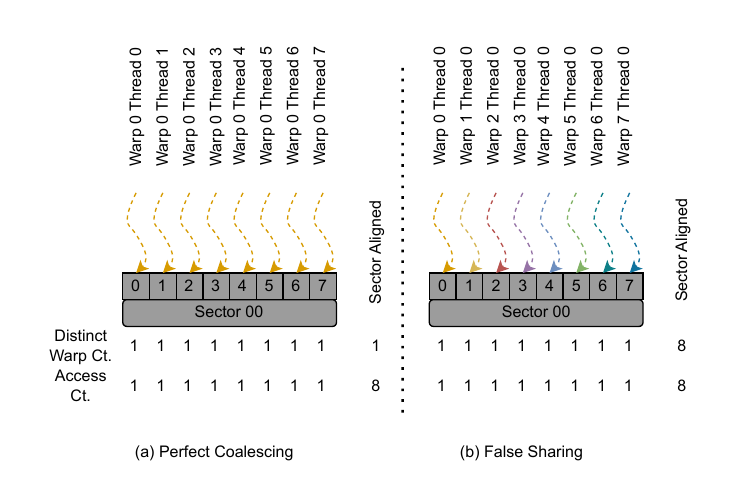}
    \vspace{-12pt}
    \caption{Perfect coalescing vs. false sharing patterns in a \gmem{} sector. Our heat map can distinguish these two patterns, whereas the memory access counting method cannot.}
    \label{fig:false_sharing}
    \vspace{-8pt}
\end{figure}

Since memory instructions are issued at the warp level, as noted in \Cref{section:background}, the number of distinct warps accessing a memory region serves as a reliable indicator of data sharing within an SM.
\Cref{fig:false_sharing} illustrates the distinct warp count for perfect coalescing and false sharing patterns, respectively. In \Cref{fig:false_sharing}(a), the words are accessed by threads within a single warp, forming a perfect coalescing pattern, resulting in a distinct warp count of 1. Conversely, in \Cref{fig:false_sharing}(b), where the count is 8, eight distinct warps access different words in the sector, representing a false sharing pattern.
The distinct warp count metric effectively describes the data-sharing level because it directly reflects the number of unique warp contributors, capturing the diversity of access patterns.
% Additionally, it accounts for potential conflicts or inefficiencies, that simple access counts overlook.

% In optimizing memory resource usage within an SM, the most reliable indicator of data sharing among warps is the number of distinct warps accessing a memory region, as illustrated in \Cref{fig:false_sharing}. 
% memory instructions are issued at the warp level, and as noted in \Cref{section:background},
% For instance, the Warp Access Count alone fails to distinguish sharing levels across sectors. In \Cref{fig:false_sharing}(a), a single warp accesses Sector with \textsc{memory access count} of 8, but no other warps share this region, indicating perfect coalescing. In contrast, \Cref{fig:false_sharing}(b) shows multiple warps accessing the same sector with the same \textsc{memory access count} of 8, yet the Distinct Warp Access Count of 8 reveals false sharing. This highlights that relying solely on access frequency misrepresents the sharing level. \textsc{Distinct Warp Count} effectively describes sharing level because it directly reflects the number of unique warp contributors, capturing the diversity of access patterns. Additionally, it accounts for potential conflicts or inefficiencies, such as false sharing, that simple access counts overlook.

% \subsubsection{Choosing Granularity}
\subsubsection{Granularity---Word-Sector}
% Choosing of granularity can significantly influence the detection of inefficiency patterns. An important thing about memory instruction processing is when a memory instruction issued from subcore to LD/ST unit, the coalescing process will merge all required memory region into sector aliased requests. So the data locality in one issued instruction will be directly responsible to the actual memory transaction. In Fig~\ref{fig:false_sharing}a, the perfect coalesced memory request from same warp will only cause $1$ sector transaction. But false-sharing in Fig\ref{fig:false_sharing}b will lead to $8$ sectors memory transaction, which can not be observed by memory count in any levels. In word level, the only observation is the element is not shared or not; and in sector-level, how this entire sector been shared is not available. A better solution which we deploy in \tool{} is combining the word sharing level and sector sharing level, then the reason of why redundant memory transaction been introduced will be able to been easily observed. 
% false sharing of a sector between warps. Access count can not detect False-sharing because it doesn't account warp id information.
% And after all trace been analyzed, the amount of $1$s will be the number of distinct warps who touched this sector. And for better observation of in sector's memory sharing, we deploy this bitmask measurement in word($4\ bytes$) level too. And group them in a sector aligned layout to display inner sector pattern and inter sectors pattern. 

The choice of granularity significantly impacts the detection of inefficiency patterns in memory operations.
At the word-level granularity in the memory access counts method, analysis is restricted to determining whether an element is shared, offering limited insight. Sectors serve as the actual memory transaction units after hardware memory request coalescing.
When a memory instruction is issued from a subcore to the LD/ST unit, the coalescing process consolidates targeted memory regions into sector-aligned requests, making data locality within a single instruction a key factor in determining actual memory transactions. As shown in \Cref{fig:false_sharing}(a), a perfectly coalesced request from a single warp results in only one sector transaction. In contrast, false sharing in \Cref{fig:false_sharing}(b) triggers eight sector transactions.
However, if only the sector level is considered, the data sharing between words within a sector remains untraceable. To overcome these shortcomings, we record distinct warp counts at both the word and sector levels, enabling precise identification of the causes behind redundant memory transactions.

\subsection{\tool{} Implementation}\label{Sec:Implementation}
This section describes the three main components of \tool{}: Trace Collector, Analyzer, and GUI, which work together to construct and visualize the heat map, using distinct warp count as the metric and word-sector granularity to reveal memory behavior and identify memory inefficiencies. To support other GPU platforms, developers only need to modify Trace Collector slightly to implement platform-specific instrumentation, and reuse other components.

\subsubsection{Trace Collector}
The Trace Collector consists of two main components: binary instrumentation and the trace packer.
In the binary instrumentation phase, \tool{} captures the user's execution command line and isolates the cubins from the target executable. It then parses these cubins with \texttt{NVBit} to inject our collection function at the appropriate locations by selecting the correct \texttt{op} code type (memory \texttt{op}s in \tool{}).
% \todo{What op code type? depending on what kind of instruction we want to instrumentation. In this work it is memory instructions.}
This allows us to capture the associated attributes: \texttt{pc}, \texttt{address}, \texttt{size}, \texttt{active\_mask}, \texttt{access\_flags}, \texttt{warp\_id}, and \texttt{block\_id}, for all issued memory instructions. A description of these fields is provided below:
% The \texttt{Trace Collector} contains two main part, binary instrumentation and trace packer. The \tool{} captures user's execution command line and the isolate cubins from target executable. Then parse those cubins with support of \texttt{NVBit} to inject our collection function into those cubins right place, by selection of right \texttt{op} code type. We can capture all memory instructions with its \texttt{address, }\texttt{size, }\texttt{active\_mask, }\texttt{flags,} \texttt{warp\_id} and \texttt{block\_id}. The describe of those fields like below:
\begin{enumerate}
    \item \texttt{pc}: Program counter.
    \item \texttt{address}: A 32-element array of unsigned integers, recording all target addresses (global memory) accessed by this issued instruction within the warp.
    \item \texttt{size}: The required length of the target address.
    % \item \texttt{pc}: Program counter, identifies instruction in executable.
    \item \texttt{active\_mask}: Indicates active threads in the warp.
    \item \texttt{access\_flags}: Specify the access type and scope (e.g., load/store/atomic and shared/local/global).
    \item \texttt{warp\_id}: Warp index within a block, from $0$ to $31$.
    \item \texttt{block\_id}: Block index, used for block sampling.
    % \item \texttt{address}: 32 elements array of unsigned int, recording all target addresses(global memory) from this issued instruction of this warp.
    % \item \texttt{size}: The required length of target address.
    % % \item \texttt{pc}: Program counter, identifies instruction in executable.
    % \item \texttt{active\_mask}: Which thread in this warp will be active.
    % \item \texttt{flags}: The access type and scope. Shared/Local/Global, Load/Store/Atomic.
    % \item \texttt{warp\_id}: The warp id in block, from $0$ to $31$.
    % \item \texttt{block\_id}: The block id. For block sampling.
\end{enumerate}

In the trace packer, a \texttt{GPU queue} is created, containing a buffer and synchronizing flags to facilitate information transfer between the CPU and GPU.
\tool{} packs the seven attributes for all issued memory instructions into the buffer and waits for the Analyzer to copy them to CPU memory as its input. \textit{Trace Collector} also provides callback functions responsible for handling CUDA events, such as \texttt{cudaMalloc} (for global memory), \texttt{cudaMallocManaged} (for unified memory), \texttt{kernel launch}, and \texttt{kernel end}, to assist the Analyzer in organizing the collected trace.

% By using \texttt{GPU queue} with sync flags between CPU and GPU, \tool{} can pack those collected information into a buffer and waiting for the Analyzer to copy them back to CPU memory as \textit{Analyzer}'s input. And provides callback functions who will been responsible for CUDA events, like \texttt{cudaMalloc(global memory)}, \texttt{cudaMallocManaged(Unified memory)}, \texttt{kernel launch} and \texttt{kernel end} to help the \texttt{Analyzer} organize the collected trace.

% \jli{Too short for sampling, this is one contribution of the work. Explain why we need to do sampling.}
% The block-sampling policy is implemented here. The injection function contains a bypass check to make sure only target block's trace will been collected. Trace from other blocks will been ignored to reduce collection overhead and noise. 

\noindent \textbf{Thread Block-Sampling.}
However, the trace size of a GPU application, even when limited to memory instructions, is too large to be stored entirely in GPU memory.
More importantly, since identical warp IDs are reused across different blocks, this can lead to misinterpretation of memory access patterns, inefficiencies, and even make trace-based pattern detection impossible.
To address this, we propose a thread block-sampling approach that collects the trace from only a single target thread block by including a bypass check in the injection function.
Although simple, this method not only reduces collection overhead and thus makes \tool{} applicable to large applications, but also ensures accurate detection of memory access patterns by minimizing noise from other warps in different thread blocks.

% The block-sampling policy is implemented here to efficiently manage trace collection in GPU memory optimization. The injection function includes a bypass check to ensure that only traces from target blocks are collected, discarding those from other blocks to minimize collection overhead and reduce noise. This targeted approach is crucial because, when observing behavior within a target block, traces from other blocks can distort the analysis. This interference arises primarily from warp ID reuse across blocks, where identical warp IDs from different blocks may overlap, leading to misinterpretation of access patterns and inefficiencies. this method not only reduces computational overhead but also enhances the precision of detecting redundant memory transactions.

\subsubsection{Analyzer}
The Analyzer maintains the status of synchronization flags in the \texttt{GPU Queue} and is responsible for transferring collected information between the CPU and GPU, then analyzing the collected traces to generate configuration files and CSV files that record heat maps.
When a \texttt{GPU Queue} buffer is full, the Analyzer copies its contents to CPU memory and invokes the analyzing function to process the data.
Additionally, the Analyzer implements memory registration functions to store memory allocation information captured by the Trace Collector. 
% The Analyzer maintenance the \texttt{GPU Queue}'s synchronization flags between CPU and GPU, and responsible to do analyze on those collected traces. When a \texttt{GPU Queue}'s buffer is full, the \texttt{Analyzer} will copy them to CPU memory and call analyzing function to perform our analyzing process. Besides them, the analyzer implement memory registration functions to store the memory allocation information captured by \texttt{Trace Collector}. 

The Analyzer maintains a map, \texttt{sector\_history\_map}, using sector tags as keys and a size-$9$ array, \texttt{size\_t[9]}, to record the history of distinct warp accesses. Every element is the bitmask to record the warp IDs into corresponding positions. The first eight elements of the value array represent each 4-byte \texttt{word} within the sector, while the last element represents the entire sector.
% Analyzer maintenance a map, \texttt{sector\_history\_map}, using \texttt{sector tag} as the key, \texttt{uint64\_t[9]} as value whose first 8 elements represent every 4-byte \texttt{word}, last element for the entire sector's bitmasks to record how many distinct warp access history. 

The analyzing function accepts a trace buffer grouped by warps, where each buffer element contains the aforementioned seven attributes, to fill in the \texttt{sector\_history\_map}. All the information other than \texttt{address} is shared over the threads within a warp when a memory instruction is issued. 
The addresses are then processed to obtain the \texttt{sector tags}, each of which denotes the starting address of a sector. Since the sector size is $32$ bytes, the tag is calculated as $\texttt{sector tag} = \texttt{address} / 32$ and the word offset within a sector as $\texttt{offset} = \texttt{address} \% 32$.
The corresponding entry in \texttt{sector\_history\_map} is updated by using $1 << \texttt{warp\_id}$ and applying the bitwise OR ($\texttt{|=}$) operation with the bitmask in \texttt{sector\_history\_map[tag][offset]} and \texttt{sector\_history\_map\allowbreak [tag][8]}, thereby storing all accessed warp IDs for all the eight words and the entire sector.
% The analyzing function accepts a trace buffer grouped by warps, every buffer element contains \texttt{pc, } \texttt{address, }\texttt{size, }\texttt{active\_mask, }\texttt{flags,} \texttt{warp\_id} and \texttt{Block\_id}. All other information is shared inside a warp when a memory instruction is issued. Then the addresses will been processed to get the \texttt{sector tag}, because the sector's size is $32$ bytes, so the $tag=addr / 32$, $offset=addr \% 32$. Then update the corresponding entry in \texttt{Sector\_history\_map}, use $1<<warp\_id$ to $|=$ with the bitmasks in \texttt{Sector\_history\_map[tag][offset]} and \texttt{Sector\_history\_map[tag][8]} to store all accessed warp id for the entire sector.

% \yanbo{Remove it to avoid repeats.}
% \jli{Put it in a proper place.}
% To measure the sharing level of a sector among warps from one block, we deploy $32$-bits \textit{Warp\_Bitmask} to record the warp ids because the maximum block size is $1024$ representing the maximum warp id is $32$ in one block. When a memory access happens, the bitmask will use $|=$ operator with $1<<warp\_id$. The \textit{Warp\_Bitmask} will store all warp ids who accessed this memory region which can perfectly reflect the sharing level among warps in a thread block. 

After all traces of a kernel have been analyzed and kernel execution has ended, a flush function is invoked to count the number of $1$s in the bitmasks of every entry in \texttt{sector\_history\_map}. These $1$s represent the number of distinct warps that have accessed each word and each sector, and they are recorded in CSV files.
Next, the start and end addresses of memory regions, along with the start and end sector tags, are stored in configuration files. 
% Using these files, the GUI component could easily display the heat map.
% After all trace of a kernel been analyzed and kernel execution is end, a flush function will been invoked to account $1$s from the bitmasks in every entry in \texttt{Sector\_history\_map} which represent the amount of distinct warps has accessed every word and every sector to generate heatmap csv files. Then store the memory regions' start address and end address and sector tags' start and end to config files. With those files, our GUI can display the heatmap more straightforwardly. 

\subsubsection{Heat Map Visualization.}

% Based on the configuration and CSV files generated by the Analyzer, \tool{} produces heat maps for better visualization.
% CSV files are used to create the heat map, while configuration files are used to determine which memory region each sector belongs to.
Based on the configuration and CSV files generated by the Analyzer, \tool{} produces heat maps to visually represent memory access patterns. The CSV files provide the raw data for constructing the heat map, while configuration files map each memory sector to its corresponding region, allowing users to distinguish between different access patterns across regions.

% \Cref{fig:gemv_heatmap} illustrates our heat map design. The horizontal axis represents sector tags, while the vertical axis corresponds to word IDs, indexing each word within a sector. The values are represented as \textit{temperature}, indicating the number of distinct warps accessing a sector—a higher temperature signifies more distinct warps.
% An extra cell at the top of each sector represents the temperature of the entire sector, which is the total number of distinct warps that have accessed it. A color-coded legend on the right side of the heat map uses a gradient of hues to represent sector temperatures.\footnote{The applications we tested set the thread block size to $256$, meaning the highest temperature is $8$, corresponding to up to $8$ distinct warps per thread block.}
% To optimize visualization, when consecutive rows in the CSV file contain identical values, the heat map generator compresses them into a single summarized entry indicating the frequency of occurrence and retains only the boundaries of repeated sector access patterns. This compression enhances readability and reduces redundancy.

% \tool{} utilizes heat maps to visualize the CSV files generated by the Analyzer. In this visualization, we use the term \textit{temperature} to represent the number of distinct warps accessing a given sector. A higher temperature corresponds to a greater number of warps accessing the sector. 

\begin{figure}
    \centering
    \includegraphics[width=0.9\linewidth]{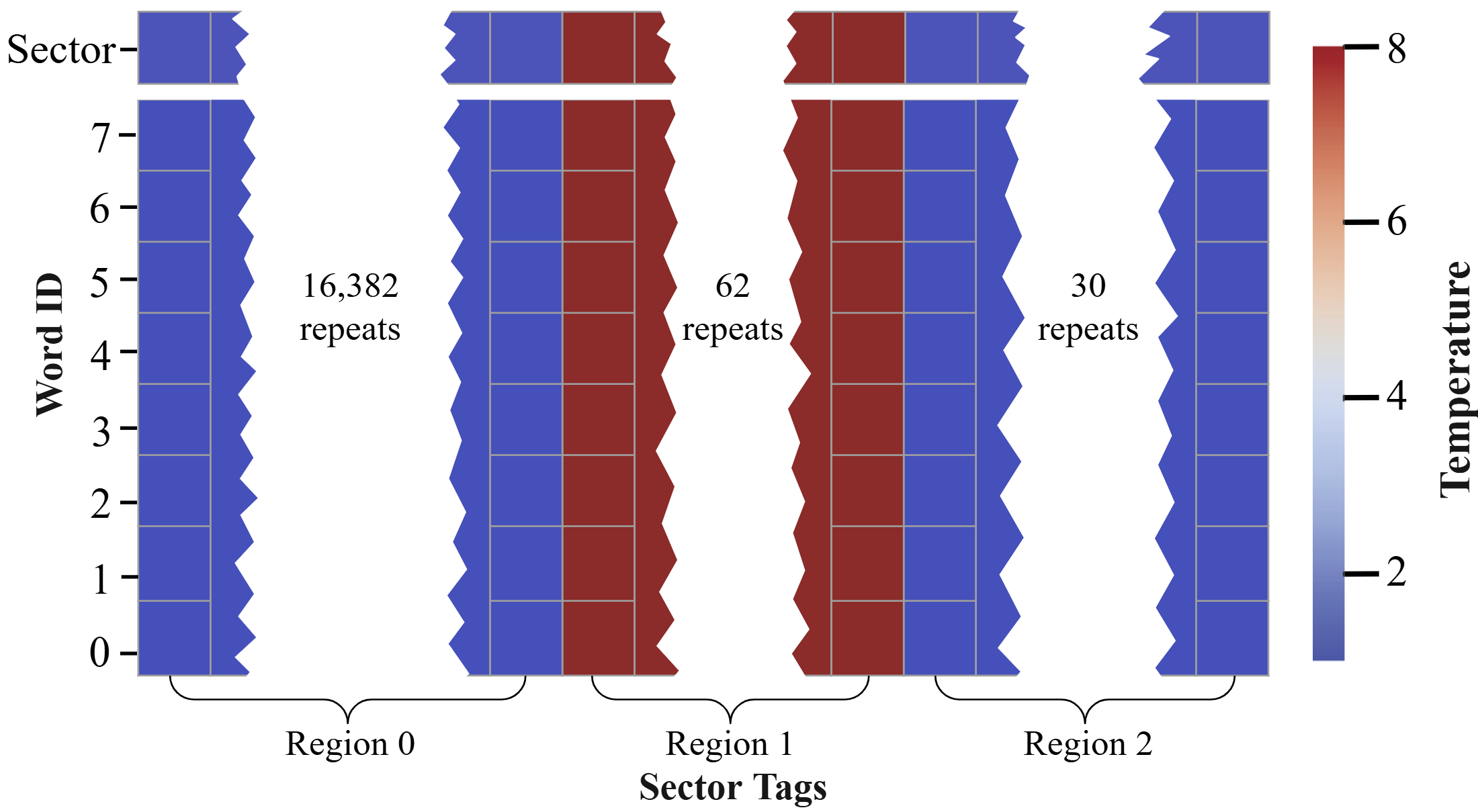}
    \vspace{-8pt}
    \caption{Heat map using region indices and sector tags. Consecutive memory regions with identical temperatures are compressed, and the number of occurrences is indicated.}
    \label{fig:gemv_heatmap}
    \vspace{-14pt}
\end{figure}

\Cref{fig:gemv_heatmap} illustrates the heat map design, where memory access patterns are categorized based on \textit{region indices} while retaining sector tags for detailed access tracking. The horizontal axis represents sector tags, while the vertical axis corresponds to word IDs, indexing each word within a sector. The color intensity in each cell represents the \textit{temperature}, indicating the number of distinct warps accessing a given sector—a higher temperature reflects greater memory sharing. An additional cell at the top of each sector represents the aggregated temperature of the entire sector, showing the total number of distinct warps that have accessed it. To enhance readability, consecutive memory regions with identical access patterns are grouped and labeled with their repetition count. A color-coded legend on the right uses a gradient of hues to visualize sector temperatures.

The heat map provides an intuitive representation of memory access intensity, helping users identify critical hot regions such as Region~1 in \Cref{fig:gemv_heatmap}. These regions, characterized by high temperatures, indicate highly shared sectors, where optimizing data placement in registers, shared memory, or L1 cache could improve performance. Conversely, cold regions with minimal access frequency suggest a potential streaming memory access pattern, where bypassing cache accesses may help prevent unnecessary cache pollution.

\begin{figure}[htbp]
    \centering
    \includegraphics[width=0.7\linewidth]{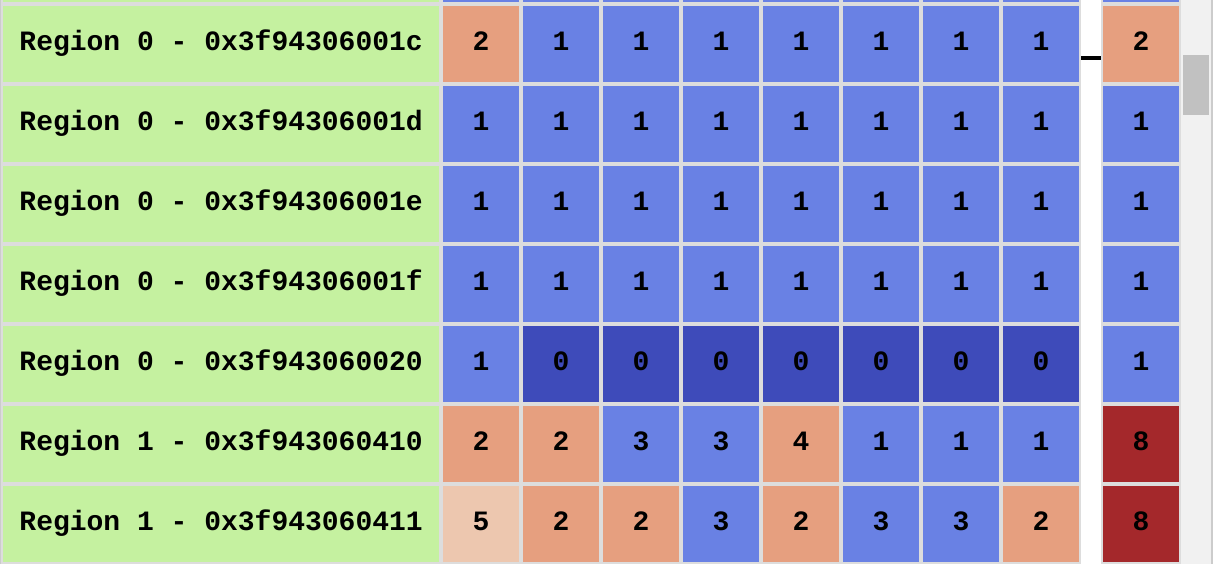}
    \caption{Implementation of the heat map in \tool{}, organized in a vertical layout for enhanced usability.}
    \label{fig:spmv-heatmap}
    \vspace{-1em}
\end{figure}

\Cref{fig:spmv-heatmap} presents a screenshot of the heat map in \tool{}'s user interface. The visualization adopts a \textit{vertical layout} to improve navigation and facilitate pattern recognition. The vertical scrolling feature allows users to scan through the memory access patterns efficiently, making it easier to pinpoint optimization opportunities at a glance.

By offering a clear and structured heat map visualization, \tool{} can identify memory access patterns and assist users in applying appropriate optimization strategies.

\begin{figure}[h]
    \vspace{-0.45cm}
    \centering
    \includegraphics[width=\linewidth]{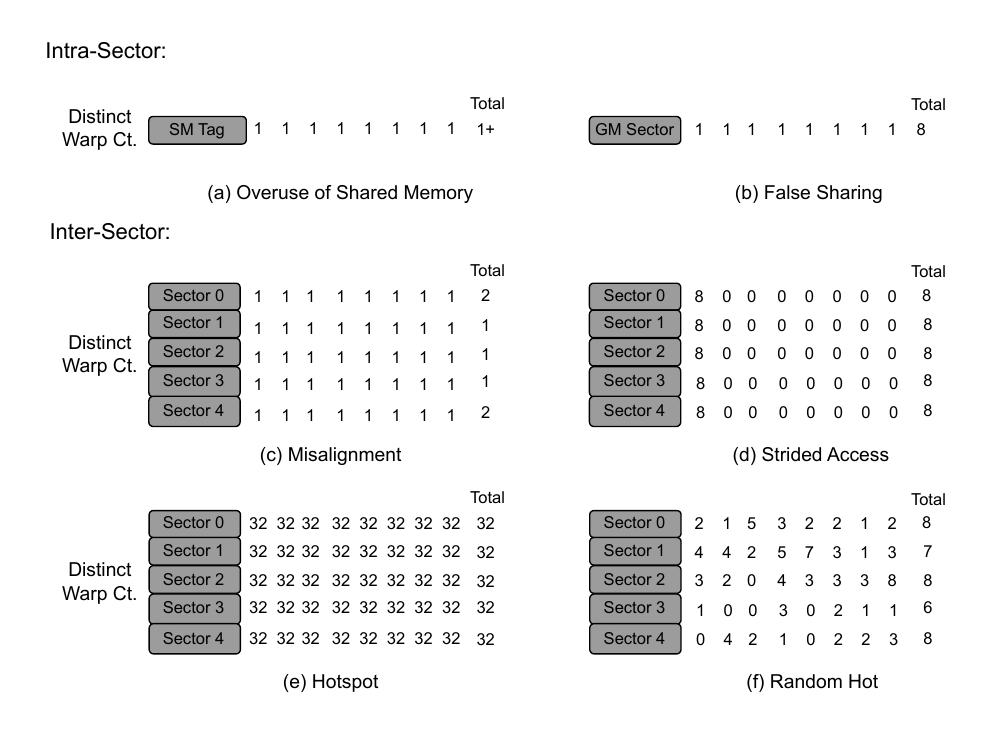}
    \vspace{-1cm}
    \caption{Identified memory access patterns via \tool{}.}
    \label{fig:patterns}
    \vspace{-0.5cm}
\end{figure}
\subsection{Identified Patterns}
\label{sec:patterns}

We showcase five memory access patterns identified by \tool{} that are either impossible or difficult to detect using state-of-the-art methods. These patterns are presented as example heat maps in \Cref{fig:patterns}, with their underlying causes and corresponding solutions explained in this section.
These patterns are not limited to a specific NVIDIA GPU generation; they are broadly applicable across different NVIDIA GPU architectures. For optimization guidance, we provide detailed examples in \Cref{sec:cases}.  

\noindent \textbf{(Random) Hot Spots. } Hot sectors are identified by high temperatures in each word of certain sectors, and the sector temperature is close to the highest in its words. \Cref{fig:patterns}(e) is a typical example of hot sectors which are accessed by 32 warps. \Cref{fig:patterns}(f) is a case of random hot pattern, which shows random numbers of warps sharing the words. 

\noindent \textbf{Abuse of Shared Memory. }
Since \smem{} is configurable and physically co-located with the L1 cache, incorrect use of \smem{} can significantly impact the performance of the L1 cache and other hardware components due to the reduction in L1 cache capacity and the introduction of extra memory or synchronization instructions.
% The \smem{} is located in the same place with the L1 cache physically, which has configurable size in modern GPUs. But incorrect use of \smem{} might significantly influent the performance of L1 cache and SM because of reduce of L1 cache capacity and extra memory/sync instructions introduced. 
\Cref{fig:patterns}(a) gives the pattern of one line of our heat map corresponding to the particular sector tag that belongs to the \smem{} address space.
% and the sector belongs to \smem{} address space.
% This pattern describes an array resided in the \smem{} space continuously accessed by threads in a warp.
Every word's temperature in this \smem{} tag is 1, while the number of distinct warps accessing this entire 32-byte region is 1 warp (as shown in this figure) or multiple warps depending on application scenarios. 
That means each warp is accessing different words in \smem{} during execution, which is against the purpose of using \smem{}: to shorten access latency for data reused multiple times~\cite{sharedmemroy}.
% The overuse of the precious \smem{} resources could influence the performance of the entire application by reducing L1 cache size.
Overusing precious \smem{} resources can impact the overall application performance by reducing the available L1 cache size.

% \emph{Solution: } 
% Use substantial registers instead of \smem{}. If communication within a warp is required for this data region, it can be performed using shuffle instructions.
% Use substantial registers instead of \smem{}. If any communication within a warp is needed in this data region, it can be performed using shuffle instructions.

% If a shared memory region will only be accessed within one warp.  
% To detect this pattern, we can check the sector tag's belonging and the sector temperature. 
% If every 4-byte word's temperature in a sector is 1 and the sector belongs to Shared Memory address space.
% It can be concluded as abuse of shared memory. Because there will only be one warp accesses this memory region during exection, the communication inside warp can be performed by \textbf{shfl} instructions instead of using expensive shared memory resources.

\noindent \textbf{Memory False Sharing. }
A warp performs a hardware coalescing process that merges memory access instructions from the same warp into sector-aligned memory transactions, after instructions have been issued. 
Thus, the memory access requests to different words of a sector will only cause 1 actual sector transaction within one warp. 
However, memory access instructions from multiple warps cannot take advantage of coalescing, thus leading to 8 sector transactions.
This pattern of false sharing~\cite{Will1993} on \gmem{} limits the application performance.
Our heat map identifies this pattern, shown in \Cref{fig:patterns}(b), where each word is accessed by one distinct warp, triggering multiple sector transactions from \gmem{}. The false sharing pattern differentiates from the hot spot pattern by the total number of distinct warps accessing an entire sector versus those accessing individual words within that sector. When false sharing occurs, the whole sector's temperature is significantly higher than any single word within that sector. 

% \emph{Solution:} False sharing could be avoided by reorganizing data. For example, when a matrix is accessed in column-major order in one kernel but row-major in another, an explicit transpose operation can make both access patterns use row-major to reduce false sharing, improving efficiency. 

% Or memory instruction from multiple warps to the same sector will lead to extra memory requests. 
% The false sharing pattern can be detected by our heatmap design in this way: A sector have a similar distinct-warp count in its 4-bytes words align view and a significant higher number in sector aligned view. Which means every warp will only use a little part of this sector.

% As we discribe in section~\ref{section:background}, a warp will perform a coalescing process after memory instruction has been issued and before the memory request has been assigned into LD/ST unit. Which will merge load instructions from same warp into a sector aligned memory requests. But memory request from other warps can not directly take advantage of coalescing. Which means many memory instruction from one warp to the same sector will only cause 1 sector actual transaction. Or memory instruction from multiple warps to the same sector will lead to extra memory requests. The false sharing pattern can be detected by our heatmap design in this way: A sector have a similar distinct-warp count in its 4-bytes words align view and a significant higher number in sector aligned view. Which means every warp will only use a little part of this sector.

\begin{figure}[h]
    \centering
    \vspace{-10pt}
    \includegraphics[width=1\linewidth]{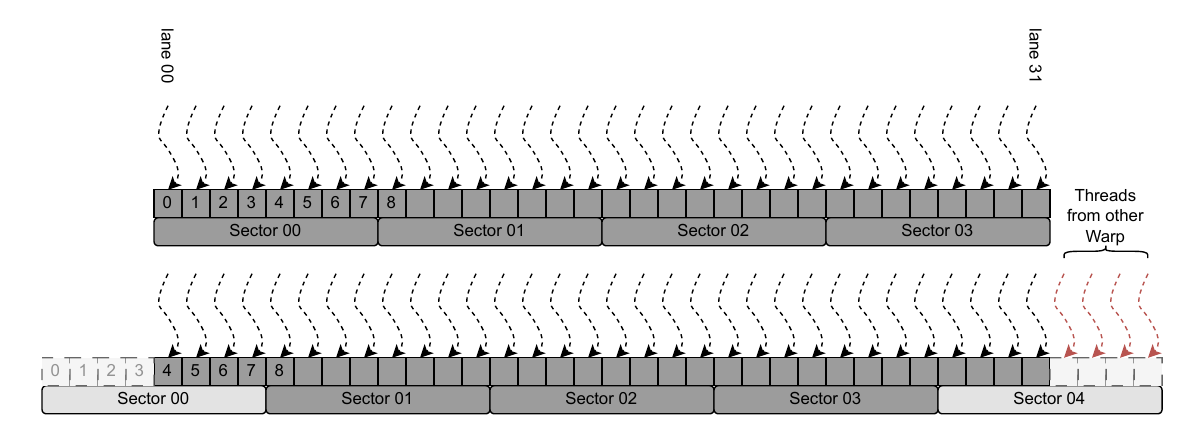}
    \vspace{-12pt}
    \caption{Misaligned memory request. Five sectors need to be loaded into the L1 cache for warp 1, but the first and last four elements in sectors 0 and 4 will never be used.}
    \label{fig:sector_misalignment}
    \vspace{-6pt}
\end{figure}
% Misaligned memory request, if current warp is warp 0, array[0:3] and array[32:35] will never been used but the entire sector 00,04 will be fully loaded into L1 cache. \jli{Figure too small, hard to read. Use 8 threads to represent, not 32.

\noindent \textbf{Memory Misalignment. }
Memory requests must be aligned to 32-byte sectors, as the smallest granularity, to be read or written via memory transactions to \gmem{}.
\Cref{fig:sector_misalignment} shows the misalignment pattern from our heat map measurement using five example sectors.
The internal three sectors 1-3 in a 128-byte memory region show normal behavior where the threads in one warp access different words of a sector.
However, the two boundary sectors 0 and 4 have two warps accessing words 4-7 and words 0-3 respectively, leading to one more sector transaction (5 vs. 4) from global memory to L1 cache and a waste of L1 cache due to semi-utilized sectors.
Even though modern GPUs have powerful hardware components~\cite{KernelProfilingGuide} that can handle irregular memory access patterns, those patterns still cause significant increases in memory request amounts and result in the inability to capture benefits by using built-in vectorized types.
% And will lead to more L1 cache pressure cause the semi-utilized sectors will still take an entire sector in L1.
% like MIO queue
However, the extra sector transaction cannot be revealed from memory access counting because the memory access is always 8 for a sector.

\noindent \textbf{Strided Memory Access. }
Strided memory access refers to a memory access pattern in which consecutive threads access memory locations that are spaced apart by a fixed stride, rather than accessing contiguous locations.
As shown in \Cref{fig:patterns}(d), eight warps access the same word location across five sectors, with a stride size of $7$. Seven out of eight words in each sector are never accessed by any warp during execution, meaning that $7/8$ of the memory transactions are wasted.
Additionally, the accessed words record a distinct warp count of 8, indicating that memory coalescing does not apply. This strided access pattern also wastes L1 cache due to a large number of unused words, reducing the available L1 cache for other data objects.

\section{Evaluation}\label{Sec:Evaluation}

This section evaluates our work from three perspectives: the patterns identified in real applications and benchmarks via \tool{}, as well as the tool's overhead and portability.

% \textbf{Platform specification:}\label{platform spec}:
We evaluate \tool{} on an x86\_64 system equipped with an AMD Ryzen Threadripper PRO 5955WX 16-core processor and an NVIDIA Ada Lovelace RTX4090 GPU. The following system software is used: Linux 5.15.0-131-generic, NVIDIA CUDA Toolkit 12.1.r12.1, NVIDIA Driver 535.183.01, NVBit 1.7.3, and GCC 11.4.0. 

% \jli{2. low-overhead for users and low runtime overhead, no recompiling, no LLVM passes. }

\begin{table}[htbp]
\centering
\vspace{-6pt}
\caption{Representative memory access patterns detected by \tool{} in five applications across multiple domains. }

\label{tab:access-patterns}
\resizebox{\linewidth}{!}{
\begin{tabular}{c|c|c|c|c}
\toprule
\textbf{Domain}                                                               & \textbf{Application}               & \textbf{CUDA Kernel}                                                 & \textbf{Data Object} & \textbf{Access Pattern}                                                          \\ \toprule
\multirow{6}{*}{Linear Algebra}                                               & \multirow{4}{*}{\textbf{GEMM}}     & \multirow{3}{*}{gemm\_v00}                                           & A                    & Hot                                                                              \\ \cline{4-5} 
                                                                              &                                    &                                                                      & B                    & False shared                                                                     \\ \cline{4-5} 
                                                                              &                                    &                                                                      & C                    & False shared                                                                     \\ \cline{3-5} 
                                                                              &                                    & gemm\_v01                                                            & B                    & Hot                                                                              \\ \cline{2-5} 
                                                                              & \multirow{2}{*}{\textbf{SpMV}}     & \multirow{2}{*}{spmv\_csr}                                           & rowOffsets           & Misaligned                                                                       \\ \cline{4-5} 
                                                                              &                                    &                                                                      & x                    & Hot-random                                                                       \\ \midrule
\begin{tabular}[c]{@{}l@{}}Tensor Algebra\end{tabular}                     & \textbf{PASTA}                     & \begin{tabular}[c]{@{}l@{}}spt\_TTMRankRBNnzKernelSM\end{tabular} & Y\_shr               & Abused SMEM                                                                    \\ \midrule
\multirow{3}{*}{Solver}                                                       & \multirow{3}{*}{\textbf{GRAMSCHM}} & \multirow{2}{*}{kernel2}                                             & q                    & Strided                                                                          \\ \cline{4-5} 
                                                                              &                                    &                                                                      & r                    & Strided                                                                          \\ \cline{3-5} 
                                                                              &                                    & kernel3                                                              & q                    & Strided, hot                                                                     \\ \midrule
\multirow{2}{*}{Compression}                                                  & \multirow{2}{*}{\textbf{cuSZp}}    & \multirow{2}{*}{cuSZp\_(de)compress\_kernel\_*}                      & exel\_sum            & \multirow{2}{*}{Abused SMEM}                                                   \\ \cline{4-4}
                                                                              &                                    &                                                                      & base\_idx            &                                                                                  \\ \midrule
\multirow{3}{*}{\begin{tabular}[c]{@{}l@{}}Molecular\\ Dynamics\end{tabular}} & \multirow{3}{*}{\textbf{GPUMD}}    & find\_cell\_counts                                                   & cell\_count          & \multirow{3}{*}{\begin{tabular}[c]{@{}l@{}}Strided,\\ false shared\end{tabular}} \\ \cline{3-4}
                                                                              &                                    & \multirow{2}{*}{find\_cell\_contents}                               & cell\_count          &                                                                                  \\ \cline{4-4}
                                                                              &                                    &                                                                      & cell\_count\_sum     &                                                                                  \\ \bottomrule
\end{tabular}

}
\vspace{-1em}
\end{table}

% We select kernels from those applications with representative patterns.

\noindent \textbf{Patterns in Applications.}
We use \tool{} to detect the patterns described in \Cref{sec:patterns} within real applications and benchmark suites across five domains.
The memory access patterns and their corresponding data objects from six applications and their kernels are listed in \Cref{tab:access-patterns} as representatives.  

General Matrix-Matrix Multiplication (\textbf{GEMM}), a fundamental operation in linear algebra, is widely used in scientific computing and machine/deep learning.  
Its kernels, \texttt{gemm\_v00} and \texttt{gemm\_v01} \cite{cuda-gemm-opt}, exhibit hot spots and false sharing patterns.  
Sparse Matrix-Vector Multiplication (\textbf{SpMV})~\cite{spmv} is a sparse linear algebra operation that efficiently processes matrices with mostly zero elements by storing only nonzero values in compressed formats.  
The \texttt{spmv\_csr} kernel exhibits a misalignment pattern when accessing row pointers (\texttt{rowOffsets}) and a hot but random memory access pattern in vector \texttt{x}.  
\textbf{PASTA}~\cite{pasta}, a Parallel Sparse Tensor Algorithm benchmark suite, is designed to evaluate sparse tensor algebra on different computer systems.  
The kernel \texttt{spt\_TTMRankRBNnzKernelSM} exhibits a shared memory abuse pattern on \texttt{Y\_shr} object.  
\textbf{GRAMSCHM} is a benchmark obtained from the PolybenchGPU suite~\cite{polybench}, implementing the Gram-Schmidt algorithm on GPUs.  
The \texttt{kernel3} exhibits strided memory access and hot spot patterns on the data object \texttt{q}.  
\textbf{cuSZp}~\cite{cuszp} is an ultra-fast, error-bounded GPU lossy compressor.  
The two data objects, \texttt{exel\_sum} and \texttt{base\_idx}, in kernels prefixed with \texttt{cuSZp\_(de)compress\_\allowbreak kernel\_} exhibit shared memory abuse pattern.  
\textbf{GPUMD}~\cite{gpumd} is a highly efficient, general-purpose molecular dynamics (MD) package widely used in the physics and chemistry communities.  
The kernels \texttt{find\_cell\_counts} and \texttt{find\_cell\_contents} exhibit strided and false sharing patterns in \texttt{cell\_count} and \texttt{cell\_count\_sum}.  
% The observed patterns indicate potential inefficiencies in these applications. We optimize these kernels based on our analysis in \Cref{sec:cases}.
The observed patterns reveal potential inefficiencies in these applications, motivating the kernel optimizations detailed in our case studies in \Cref{sec:cases}, with the resulting performance improvements summarized in \Cref{tab:speedup}.

\begin{table}[htbp]
\centering
\vspace{-1em}
\caption{Profiling overhead of \tool{} and Nsight Compute (NCU).}
\label{tab:overhead}
\resizebox{\linewidth}{!}{
\begin{tabular}{lccc|cc}
\toprule
\multirow{2}{*}{\textbf{Application}} & \multicolumn{3}{c|}{\textbf{Runtime (Seconds)}} & \multicolumn{2}{c}{\textbf{Overhead}} \\ \cline{2-6}
& \textbf{\begin{tabular}[c]{@{}c@{}}Original\end{tabular}} & \textbf{\begin{tabular}[c]{@{}c@{}}\tool{}\end{tabular}} & \textbf{\begin{tabular}[c]{@{}c@{}}NCU\end{tabular}} & \textbf{\begin{tabular}[c]{@{}c@{}}\tool{}\end{tabular}} & \textbf{\begin{tabular}[c]{@{}c@{}}NCU\end{tabular}} \\ \toprule
\textbf{GEMM}       & 0.27 & 2.61  & 1.64    & $9.86\times$  & $6.20\times$ \\ \hline
\textbf{SpMV}       & 1.19 & 1.36  & 2.08    & $1.14\times$  & $1.75\times$ \\ \hline
\textbf{PASTA}      & 3.38 & 3.62  & 5.07    & $1.07\times$  & $1.50\times$ \\ \hline
\textbf{GRAMSCHM}   & 1.35 & 77.90 & 1022.52 & $57.53\times$ & $755.19\times$ \\ \hline
\textbf{cuSZp}      & 3.83 & 7.54  & 7.99    & $1.97\times$  & $2.08\times$ \\ \hline
\textbf{GPUMD}      & 0.29 & 4.23  & 16.90   & $14.49\times$ & $57.90\times$ \\ \bottomrule
% \textbf{SpMM\_csr}  & 0.15 & 0.37  & 1.37    & $2.47\times$  & $9.19\times$ \\ 
\end{tabular}
}
\end{table}

\noindent \textbf{\tool{} Overhead.}
We evaluate and compare the performance overhead of our \tool{} (trace-based) and NVIDIA Nsight Compute (NCU, counter-based) on all the applications in \Cref{tab:overhead}.  
We measure the average end-to-end execution time over 10 runs for the original application, the application running with \tool{}, and the application profiled using NCU with all memory metrics enabled. The last two columns of \Cref{tab:overhead} report the overhead of \tool{} and NCU. 
In all studied applications, the overhead of \tool{} is significantly lower than that of NCU, because \tool{} only samples trace from one block per kernel launch, which will only generate at most 1024 threads' memory trace. \tool{} also supports kernel sampling, similar to NCU; we enabled the kernel sampling in both \tool{} and NCU in overhead comparison for fairness. The target kernel list contains the inefficient kernels we found in \Cref{tab:access-patterns}.
% The overhead of \texttt{NCU} in GRAMSCHM and GPUMD is significantly higher than the other applications, which is caused by the large number of kernel launches.
% Additionally, the GRAMSCHM and GPUMD applications run forever in both \tool{} and NCU so we do not list them in \Cref{tab:overhead}.
\footnote{The SpMV application operates on a $36,417\times36,417$ sparse matrix with $2,190,591$ nonzeros \cite{spmv-pdb1hys}, while PASTA processes a $154,985\times48,476\times76,728\times94,578$ tensor with $349,808$ nonzeros \cite{pasta-trivago-clicks}. cuSZp uses its built-in benchmark input, and GEMM executes matrix multiplication on two $1024\times1024$ square matrices. NCU use kernel replay mode (default).}

For the SpMV, PASTA, cuSZp applications, the overhead remains relatively modest, ranging from $1.07\times$ to $1.97\times$, indicating an additional execution time of $7\%$–$97\%$. However, GEMM exhibits significantly higher overheads of $9.86\times$ and $6.20\times$ for \tool{} and NCU, respectively.  
This substantial overhead arises because GEMM generates a large number of memory access requests, proportional to its computational complexity. As a result, the overhead of a trace-based tool like \tool{} increases accordingly. In contrast, NCU incurs a relatively lower overhead growth rate in GEMM because, as a hardware counter-based profiler, it collects metrics directly from hardware and drivers with a fixed overhead related to the duration of the kernel execution. And the overhead of GRAMSCHM and GPUMD is more related to the number of kernel launches rather than the memory access amounts.

% \Cref{tab:overhead} lists the performance overhead when deploying the \tool{} across four applications: \textbf{SpMV}, \textbf{PASTA}, \textbf{cuSZp}, and \textbf{GEMM}. The \textit{GRAMSCHMIDT} and \textit{GPUMD} both \tool{} and \textit{ncu} suffer from too many iterations of kernel execution, which introduced extra overhead of managing other callbacks rather than analyze itself.  It compares the \textit{origin\_runtime} (original execution time) with the \textit{tool\_runtime} (execution time with the tool) and \textit{ncu\_runtime}(profiling by ncu with all memory metrics), and calculates the resulting overhead as a multiplier. For the first three applications (\textit{SpMV}, \textit{PASTA}, and \textit{cuSZp}), the overhead is relatively modest, ranging from $1.09\times$ to $1.78\times$, indicating the tool adds between $9\%$ and $78\%$ to the execution time. However, GEMM shows a dramatic overhead of $48.37\times$, with the runtime increasing from just $0.2441$ seconds to $11.8081$ seconds when using the tool. As noted in the accompanying text, for heavy memory loaded applications like GEMM, the analyzing overhead is significantly higher because of the increasing of memory requests. Ncu shows similar but lower increasing overhead rate in GEMM because ncu collect metrics from hardware and drivers with fixed overhead. 

% \jli{Put some words from old Sec. 6.5 here in case needed.}

\begin{table}[htbp]
\centering
\vspace{-1.5em}
\caption{Performance improvement in A4500 and RTX4090 GPUs. Measured in cycles by NCU.}
\label{tab:speedup}

\resizebox{\linewidth}{!}{

\begin{tabular}{l|p{4.3cm}|c|c}
\toprule
\multirow{2}{*}{\textbf{Application}} & \multirow{2}{*}{\textbf{CUDA Kernel}} & \multicolumn{2}{c}{\textbf{Performance Improvement}} \\ \cline{3-4}
& & \multicolumn{1}{p{2cm}|}{\centering\textbf{A4500}} & \multicolumn{1}{p{2cm}}{\centering\textbf{RTX4090}} \\ \toprule
\multirow{2}{*}{\textbf{GEMM}}     & gemm\_v00                  & 721.79\%               & 682.82\%                 \\ \cline{2-4} 
                          & gemm\_v01                  & 26.07\%                & 20.27\%               \\ \midrule
\textbf{SpMV}     & spmv\_csr            & 1.85\%                 & 1.97\%               \\ \midrule
\textbf{PASTA}    & \parbox{2.2cm}{spt\_TTMRankRBNnzKernelSM} & 163.56\%               & 159.62\%               \\ \midrule
\multirow{2}{*}{\textbf{GRAMSCHM}} & kernel2              & 4.51\%                 & 9.19\%                \\ \cline{2-4} 
                          & kernel3              & 23.18\%                & 19.81\%               \\ \bottomrule
\end{tabular}
}

\end{table}

\noindent \textbf{Portability.}
\tool{} can detect memory access patterns and provide insights regardless of the NVIDIA GPU generation, as long as the sector-based memory system is used.
% \jli{We evaluate these insights by optimizing four applications on an NVIDIA Ada Lovelace RTX 4090 GPU --- the platform on which \tool{} is deployed --- demonstrating XX performance improvements. We further validate the portability of our approach on an NVIDIA Ampere RTX A4500, as shown in \Cref{tab:speedup}.}
We evaluate these insights and select four applications from studied applications on an NVIDIA Ada Lovelace RTX 4090 GPU, the one \tool{} deployed, and verify its portability on an NVIDIA Ampere RTX A4500 card in \Cref{tab:speedup}. 
\footnote{GPUMD and cuSZp are not included in the speedup evaluation because GPUMD requires domain experts for valid optimization, and the inefficient code snippet in cuSZp introduces only a $0.16\%$ increase in instructions. Detail of cuSZp is discussed in \Cref{sec:cases}.} Both GPUs show reasonable performance speedup, proving the insights we found are true inefficiencies and our optimization guidance is portable among different GPU architectures. 
% \Cref{tab:speedup} shows similar speedup in both A4500 and RTX4090 arrange from $1.98\%$ to $721.79\%$ with the same set of optimization methods. 
\Cref{tab:speedup} shows similar speedup results on both the A4500 and RTX 4090, ranging from $1.85\%$ to $721.79\%$ and $1.97\%$ to $682.82\%$ respectively using the same optimization methods.
% \textit{GPUMD} and \textit{cuSZp} are not shown in the speedup evaluation because \textit{GPUMD} needs domain experts for valid optimization, and the inefficient code snippet in \textit{cuSZp} introduces only $0.16\%$ extra instructions.  

% It is a x86\_64 system equipped with an Intel Xeon w7-2495X 24-core processor and an NVIDIA Ampere RTX A4500 GPU. The following system software is used: Ubuntu 22.04, Linux 6.8.0, NVIDIA CUDA Toolkit 12.5, NVIDIA Driver 555.42.02, NVBit 1.7.3, and GCC 12.3.0. 
% We deploy our tool in our evaluation platform: AMD Ryzen Threadripper PRO 5955WX with an NVIDIA Ada Lovelace RTX4090 GPU, and evaluated the insights we found with optimized version in another platform equipped with another architecture of GPUs (Ampere). Both platforms show reasonable speedup or benefit in profile results. Proofing the insights we found are true inefficiencies and our optimization guidance are portable among different GPU architectures.

\section{Case Study}
\label{sec:cases}
% \yanbo{The case can be 1.overuse of shared memory. 2 lack of use of shared memory. 3}

We showcase five applications that exhibit the four actionable inefficiency memory access patterns identified in \Cref{sec:patterns}. In this section, we focus on optimizing these applications and demonstrate the performance and efficiency improvements achieved through our optimizations. This highlights the importance of using \tool{} to identify inefficiencies and apply corrective measures. 
\footnote{For pattern \texttt{Hot} and \texttt{Random Hot}, the optimization suggestion is placing those memory regions into \smem, this optimization is well-known so we do not discuss it in this paper.}
However, due to effective performance optimization in real-world applications often requiring extensive domain knowledge, 
% As a result, some applications show only limited performance improvements. 
we anticipate further performance gains can be achieved when deployed by domain experts using \tool{}. 
% And for pattern \texttt{Hot} and \texttt{Random Hot}, the optimization suggestion is placing those memory regions into \smem, this optimization is well-known so we do not discuss it here.

% We study the inefficiency memory access patterns with five application from different areas. 
% And analyze the main cause comes from hardware, compiler or algorithm design. To distinguish our tool from existing GPU facilities including vendor provided tools and 3rd party provided tools, we study a several applications to show which kind of inefficiency our tool can easily detect and help fix inefficiencies.

\begin{figure}[h]
\vspace{-1em}
\begin{lstlisting}[
label={lst:gemm},
caption={gemm\_v00 kernel. Every thread will calculate one element in C.}
],
__global__ void gemm_v00(m, n, k, alpha, A, lda, B, ldb, beta, C, ldc){
  size_t C_row_idx{blockIdx.x*blockDim.x+threadIdx.x};
  size_t C_col_idx{blockIdx.y*blockDim.y+threadIdx.y};
  if (C_row_idx < m && C_col_idx < n) {
      T sum{0};
      for (size_t k_idx{0U}; k_idx < k; ++k_idx)
        sum+=A[C_row_idx*lda+k_idx]*B[k_idx*ldb+C_col_idx];
      C[C_row_idx*ldc+C_col_idx]=alpha*sum+beta*C[C_row_idx*ldc+C_col_idx];
}}
\end{lstlisting}
\vspace{-0.7cm}
\end{figure}

% __global__ void gemm_v00(size_t m, size_t n, size_t k, T alpha, T const* A, size_t lda, T const* B, size_t ldb, T beta, T* C, size_t ldc){
\subsection{GEMM - False Sharing}

General Matrix-Matrix Multiplication (GEMM) is a fundamental operation in linear algebra that multiplies two matrices to produce a third one. We study GEMM using a public GitHub repository~\cite{cuda-gemm-opt} that contains multiple widely used optimization methods. We select the kernels \texttt{gemm\_v00} and \texttt{gemm\_v01} to identify and explain the \textit{False Sharing} pattern and its optimization.  

In the \texttt{gemm\_v00} kernel (illustrated in \Cref{lst:gemm}), each thread accesses matrices \texttt{A} and \texttt{B} and updates the result matrix \texttt{C} using \texttt{C\_row\_idx} (computed as \texttt{blockIdx.x * \allowbreak blockDim.x + \allowbreak thread\allowbreak Idx.x}) and \texttt{C\_col\_idx} (computed as \texttt{blockIdx.y * \allowbreak block\allowbreak Dim.y + \allowbreak threadIdx.y}). Due to the column-major access pattern of \texttt{B}, threads in the same iteration often fetch data spanning multiple 32-byte cache sectors, while each thread utilizes only a small portion (e.g., $4$ bytes for a single-precision floating-point number). This results in significant memory waste---up to $7/8$ of the sector’s data---since adjacent threads do not fully utilize the fetched memory, even though some reuse may occur later.

The heat map generated by \tool{} reveals a perfect match with the \textit{False Sharing} pattern in matrices \texttt{B} and \texttt{C}. 
Kernel \texttt{gemm\_v01} solved this inefficiency by swapping \texttt{C\_row\_idx} and \texttt{C\_col\_idx}, enabling threads within a warp to access contiguous memory regions. This adjustment improves alignment with the 32-byte sector granularity, maximizing the use of loaded sectors. 
By reassigning thread indices, the optimization reduces wasted memory bandwidth, ensuring more efficient sector utilization. Despite \texttt{gemm\_v01} exhibiting a slightly lower L1 cache hit rate ($94.93\%$ vs. \texttt{gemm\_v00}’s $99.22\%$) with the same number of instructions, it achieves $7.21\times$ and $6.83\times$ speedups on the \textit{A4500} and \textit{RTX 4090} GPUs, respectively.

\begin{figure}[h]
\vspace{-0.5cm}
\begin{lstlisting}[
language={C++},
label={lst:pasta},
caption={A sparse tensor-times-matrix kernel from PASTA, using shared memory to accelerate temporary results.},
]
__global__ void spt_TTMRankRBNnzKernelSM(Y_val, Y_stride, Y_nnz, X_val,...){
  extern __shared__ sptValue mem_pool[];
  sptValue * const Y_shr = (sptValue *) mem_pool; 
  ...
  for(sptIndex l=0; l<num_loops_r; ++l){
    ...
    for(sptNnzIndex nl=0; nl<num_loops_nnz; ++nl){
      ...
      int y_id= tidy * Y_stride + tidx; Y_shr[y_id] = 0;
      __syncthreads();
      if(x < Y_nnz){
        ...
        for(i = inz_begin; ...){
          ...
          Y_shr[y_id] += X_val[i] * U_val[row * U_stride + r]; 
        }
        __syncthreads();
        Y_val[x * Y_stride + r] = Y_shr[y_id];
        __syncthreads();
      }}}
  ...
}
\end{lstlisting}
\vspace{-16pt}
\end{figure}
% __global__ void spt_TTMRankRBNnzKernelSM(sptValue *Y_val, sptIndex Y_stride, sptNnzIndex Y_nnz, const sptValue * __restrict__ X_val,...)  {

% use the gemm vs shared memory boosted gemm 
\subsection{PASTA - Abuse of Shared Memory (Thread Local)}
% use pasta

% In Figure~\ref{case:abuse of shared memory}, the \texttt{Y\_shr} array is a shared memory region, whose index is unique in a thread block. Which means during the execution time of this kernel, only 1 thread will access every element in \texttt{Y\_shr}. Even the compiler realized that so in ptx, the inner loop's temporal result in line $17$ is been hold in registers instead of saving to and reloading from \texttt{Y\_shr}, it can not cancel the use of shared memory because developer has explicitly declared it. So those shared memory region is still been allocated which makes redundant use of shared memory. Form our tool's perspective, the shared memory region's \yanbo{temperature} is all $1$s which represent that only one warp will touch every word in shared memory.

PASTA~\cite{pasta} is a parallel sparse tensor algorithm benchmark suite designed to help users evaluate the performance of different computer systems in sparse tensor algebra computations. We select the \texttt{spt\_TTMRankRBNnzKernelSM} kernel (illustrated in \Cref{lst:pasta}) for sparse tensor-times-matrix computation to discuss the \textit{Abuse of Shared Memory} pattern.

This CUDA kernel suffers from a major inefficiency in its use of \smem{}. The primary issue is that \texttt{Y\_shr} is used to store per-thread computation results that do not actually need to be shared, leading to unnecessary memory operations and synchronization points. Using \tool{}, we observe the \textit{Abuse of Shared Memory} pattern.  

To resolve this issue, we replace the \smem{} usage with simple register variables (e.g., \texttt{sptValue local\_sum = 0;}) to allow each thread to accumulate its results independently. This eliminates the need for \smem{} allocation and most synchronization barriers. The only necessary global memory write occurs at the end of the computation, directly updating \texttt{Y\_val}.  
This optimization achieves a $1.64\times$ and $1.60\times$ performance speedup on the \textit{A4500} and \textit{RTX 4090} GPUs, respectively.

\subsection{cuSZp - Abuse of Shared Memory (Warp Local)}
cuSZp is an ultra-fast, error-bounded GPU lossy compressor implemented in CUDA. It exhibits a similar pattern to PASTA, as all eight kernels prefixed by \texttt{cuSZp\_compress\_kernel\_} and \texttt{cuSZp\_decompress\_kernel\_} use \smem{} to perform broadcasting within a warp.  
The key difference between PASTA and cuSZp lies in the level of data sharing. PASTA has no data sharing at all, whereas cuSZp shares data among threads within a warp. The insight here is that \smem{} should not be used for in-warp data sharing.  

To resolve this issue, we replace the use of \smem{} for \texttt{exel\_sum} and \texttt{base\_idx} with registers and instead use warp-level intrinsics, such as \texttt{\_\_shfl\_sync(0xffffffff, exel\_\allowbreak sum, srcLane)}. Additionally, we remove the related synchronization function calls.  
In this case, the corresponding inefficient lines will only be executed once which makes them hard to observe via cycles, but they can be detected by executed instruction counts. Our optimization eliminates all SMEM usage in the kernels, leading to $6.44\%$ reduction in \texttt{stall\_short\_scoreboard} cycles.
% \todo{don't understand the last sentence. Remove this line? we expained in sec 5}

% \textit{cuSZp} is an ultra-fast error-bounded GPU lossy compressor implemented in CUDA, which has a similar pattern with \textit{PASTA}, all eight kernels prefixed by \texttt{cuSZp\_compress\_kernel\_} and \texttt{cuSZp\_decompress\_kernel\_} use \smem{} to perform broadcasting within a warp. The different of \textit{PASTA} and \textit{cuSZp} is the level of data sharing, \textit{PASTA} has no data sharing at all while \textit{cuSZp} shares data among threads within one warp. The insight is that the \smem{} should not been used for in-warp data sharing. 

% To fix this issue, replace the using of \smem{}, \texttt{exel\_sum} and \texttt{base\_idx}, with registers, and use \texttt{\_\_shfl\_sync(0xffffffff, exel\_\allowbreak sum, srcLane)} etc. instead, then remove the related synchronization function calls.

% In this case, the corresponding inefficient lines will only been executed for one time and hard to been observed via Nsight Compute from randomness. \todo{don't understand.}

\begin{figure}[h]
\vspace{-0.5cm}
\begin{lstlisting}[
label={lst:gramschm},
caption={A kernel from PolybenchGPU, array $q$ accessed by column major. Only $1/NJ$ part of array $q$ will be accessed.},
]
__global__ void gramschmidt_kernel3(ni,nj,a,r,q,k){
  int j = blockIdx.x * blockDim.x + threadIdx.x;
    if ((j > k) && (j < _PB_NJ)){
      r[k*NJ + j] = 0.0;
    for (int i = 0; i < _PB_NI; i++)
      r[k*NJ + j] += q[i*NJ + k] * a[i*NJ + j];
    for (int i = 0; i < _PB_NI; i++)
      a[i*NJ + j] -= q[i*NJ + k] * r[k*NJ + j];
}}    
\end{lstlisting}
\vspace{-0.5cm}
\end{figure}
% __global__ void gramschmidt_kernel3(int ni, int nj, DATA_TYPE *a, DATA_TYPE *r, DATA_TYPE *q, int k){

\subsection{GRAMSCHM - Strided Access}
% Gramschimidt
GRAMSCHM, obtained from the PolybenchGPU benchmark suite~\cite{polybench}, is a CUDA implementation of the Gram-Schmidt algorithm, which is commonly used for matrix orthogonalization. 
In \texttt{kernel3} (illustrated in \Cref{lst:gramschm}), each thread accesses elements of the \texttt{q} array in a strided pattern via \texttt{q[i*NJ+k]}, where consecutive memory accesses are separated by entire rows. This pattern prevents coalesced memory access, a critical factor for GPU performance.  

The inefficiency stems from non-coalesced loads, where a single memory transaction cannot satisfy multiple threads in a warp. Although all threads require the same \texttt{q[i*NJ+k]} element, accessing \texttt{q[(i+1)*NJ+k]} in subsequent iterations requires jumping \texttt{NJ} elements away. This strided pattern prevents loop unrolling benefits since elements are not adjacent, leading to multiple separate memory transactions and significantly reduced memory bandwidth utilization.  
Using \tool{}, we observe that only one word in each sector is accessed, and this word is shared by multiple warps, fitting the \textit{strided access} pattern.  

We fix this issue by transposing the \texttt{q} array's indices to enable coalesced memory access, allowing threads within a warp to access adjacent memory locations simultaneously. This optimization improves memory transaction efficiency by enhancing intra-warp sector sharing while reducing inter-warp conflicts.  
The results show a $1.20\times$ speedup, a $7.69\%$ reduction in register usage (from $26$ to $24$), and $20.05\%$ fewer instructions with no precision loss on RTX 4090. On A4500, the optimization achieves a $1.23\times$ speedup, $18.16\%$ fewer instructions, and the same reduction in register usage.

\begin{figure}[h]
\vspace{-0.5cm}
\begin{lstlisting}[
label={lst:spmv},
caption={SpMV implementation based on the CSR format.},
]
__global__ void spmv_kernel(numRows, rowOffsets, colIndices, values, x, y) {
  int r = blockIdx.x * blockDim.x + threadIdx.x;
  if (r < numRows) {
    float sum = 0.0f;
    for (int i=rowOffsets[r];i<rowOffsets[r+1];++i){
      int col = colIndices[i];
      sum += values[i] * x[col];
    }
    y[row] = sum;
}}
\end{lstlisting}
\vspace{-16pt}
\end{figure}
\subsection{SpMV - Misalignment}
% can use SpMV's rowOffsets example. the +1 pattern will make access not coalescable to one sector.
Sparse Matrix-Vector Multiplication (SpMV)~\cite{spmv} is a fundamental kernel in scientific computing that efficiently processes matrices with mostly zero elements by storing only nonzero values in compressed formats such as the popular CSR format, which utilizes the \texttt{rowOffsets}, \texttt{colIndices}, and \texttt{values} arrays.  

The SpMV algorithm (shown in \Cref{lst:spmv}) suffers from misaligned memory access when threads read \texttt{rowOffsets[row+1]}. Since \texttt{rowOffsets} is of type \texttt{int} ($4$ bytes), this misalignment forces each warp to load $5$ memory sectors instead of the optimal $4$, resulting in a $25\%$ overhead in memory transactions and index calculations.  

Our optimization restructures the \texttt{rowOffsets} array by introducing duplicated storage in a zigzag pattern during preprocessing, enabling vectorized memory loads via \texttt{ldg.s32.v2} instructions. This optimization achieves a $1.85\%$ and $1.97\%$ speedup with $0.29\%$ and $0.21\%$ fewer instructions on A4500 and RTX 4090, respectively.

\subsection{cuSPARSE - Closed-Source Library}
By the methodology and implementation of \tool{}, we can also apply \tool{} to closed-source libraries, and deploy our observation to understand inefficiencies without access to the source code.
% We use \tool{} to analyze multiple kernels in cuSPARSE, which are identified with the \textit{Abuse of Shared Memory}, \textit{False Sharing}, and \textit{Memory Misalignment} patterns in \texttt{matrix\_scalar\_multiply\_kernel} and \texttt{load\_balancing\_kernel} from \texttt{spmm\_csr} . 
We use \tool{} to analyze multiple cuSPARSE kernels, identifying \textit{Abuse of Shared Memory}, \textit{False Sharing}, and \textit{Memory Misalignment} patterns within \texttt{matrix\_scalar\_multiply\_kernel} and \texttt{load\_balancing\_kernel} from \texttt{spmm\_csr}.
Due to the closed source nature of cuSPARSE, we cannot provide specific code examples or optimizations, but this case shows the capacity of identifying inefficiencies in closed source libraries with \tool{}.
% However, we can still apply our methodology to analyze the memory access patterns and identify potential inefficiencies in these kernels. This case underscores the critical role and utility of \tool{} in uncovering memory inefficiencies, even within sophisticated, commercially developed libraries.

\subsection{Discussion}
\tool{}, based on instrumentation techniques, has both limitations and advantages. Its limitations include a reliance on vendor-provided binary instrumentation frameworks, which currently restricts platform compatibility, though its modular design enables easy deployment once the necessary framework components are available. Furthermore, it lacks an automatic pattern detection mechanism, as some memory inefficiencies are complex and require human judgment; instead, it offers a GUI for users to visualize memory access patterns and select from common ones. Additionally, it doesn't guarantee the selection of the most representative code block, instead allowing users to specify the block to sample and providing an interface for other profiling tools to identify representative blocks. However, a key advantage is its introduction of a novel metric for observing memory access patterns. For long-term maintainability, a lightweight alternative implementation, \texttt{cuThermo\_light}, which relies solely on the \texttt{NVBit} framework, is provided, reducing external dependencies and ensuring sustained functionality across future GPU generations.

\section{Related Work}\label{Sec:related work}

% \jli{shorten}

Detecting and optimizing GPU memory inefficiencies remains a persistent challenge throughout the lifecycle of GPU application development. Numerous profiling and optimization techniques have been introduced to assist developers in tuning their kernels by reducing memory usage, identifying redundancies, recognizing memory access patterns, and evaluating locality characteristics.
% These approaches can be broadly categorized into two groups: memory redundancy detection and memory access pattern analysis.

\subsection{Memory Redundancies}

\noindent \textit{\textbf{Instruction Redundancies}}: Several works have been proposed to identify redundant memory instructions in CPU and GPU applications. \textit{RedSPY}~\cite{wen2017redspy} and \textit{LoadSPY}~\cite{su2019redundant}, leverage vendor-provided binary rewriters, hardware performance monitoring units, and debug registers to detect redundant memory instructions in CPUs. \textit{DARSIE}~\cite{yeh2020dimensionality} eliminates conditionally redundant memory instructions at the grid, block, and warp levels, providing insights into reducing executed instruction counts. However, these methods rely on specific hardware and software support or require custom simulators based on modified ISAs, limiting their applicability to real GPUs.

\noindent\textit{\textbf{Value Redundancies}}: While tools like \textit{Gvprof}~\cite{zhou2020gvprof} and \textit{ValueExpert}~\cite{zhou2022valueexpert} identify valuable runtime redundancies based on data values, they may not capture inefficiencies rooted in algorithmic structure or memory access patterns, which is the primary focus of our work.

% \noindent\textit{\textbf{Value Redundancies}}: \textit{Gvprof}~\cite{zhou2020gvprof} and \textit{ValueExpert}~\cite{zhou2022valueexpert} analyze value-based memory redundancies that cannot be detected at compile time, providing opportunities for optimization in highly optimized applications. These tools utilize a novel method called \textit{Value-flow} analysis to track stored values across all memory regions and classify them into redundancy types such as \textit{heavy type}, \textit{single value}, and \textit{approximate values}. While these approaches provide valuable insights into value redundancies at the memory level, they primarily focus on runtime inefficiencies rather than inefficiencies stemming from algorithm design. As a result, these techniques may overlook redundant computations or inefficient data reuse strategies that originate from the underlying algorithmic structure. 

\noindent\textit{\textbf{Memory Overuse}}: \textit{DrGPUM}~\cite{lin2023drgpum} monitors GPU memory allocation and deallocation at both the CUDA runtime and third-party framework levels. It detects wasted memory regions that remain unused after allocation or deallocated long after their last access, providing effective optimization strategies for reducing memory peak usage. 
% This approach is particularly beneficial for memory-intensive GPU applications requiring efficient global memory management. 
However, due to the lack of low-level hardware observation, this method does not provide insights into kernel design optimizations or memory access behavior at the instruction level.
\subsection{Memory Access Pattern Detection}

\noindent\textit{\textbf{Reuse Distance Analysis}}: Reuse distance analysis has been widely used in GPU cache modeling to predict memory performance~\cite{nugteren2014detailed, kiani2019rdgc, arafa2019gpus, wang2016reuse, li2015locality, arafa2020fast}. These studies incorporate detailed cache hierarchy models to evaluate memory locality and reuse patterns. However, as GPU architectures evolve, new cache components are introduced, reducing the reliability of these simulation-based techniques in modern architectures.

\noindent\textit{\textbf{Coalescing Rate}}: NVIDIA Nsight Compute provides hardware counter-based metrics to measure memory coalescing rate by reporting the average number of bytes utilized per memory sector. However, this analysis is performed at the global scope, making it insufficient for identifying inefficient memory access patterns at the instruction level. 
% Consequently, developers lack guidance on how to optimize individual memory transactions to improve coalescing rate.

\noindent\textit{\textbf{Shared Memory Bank Conflict Detection}}: Gou et al. \cite{gou2013addressing}, Horga et al. \cite{horga2022symbolic} and others\cite{ferranti2019towards,boyer2008automated} proposed static and dynamic approaches for identifying bank conflicts at compile time and runtime. While these methods effectively detect potential conflicts, they do not provide actionable insights on how to restructure memory access to avoid them, limiting their practical impact on performance optimization. 

\section{Conclusions and Future Work}\label{Sec:Conclusion}
This paper presents a novel GPU program characterization method and its implementation, \tool{}, which provides a comprehensive understanding of memory access patterns in GPU-accelerated applications.
\tool{} operates on fully optimized binaries without requiring source code, driver, operating system, or hardware modifications. It features a user-friendly heat map GUI that helps users understand memory behaviors in complex kernels.  
With \tool{}, users can optimize their kernels iteratively with guidance and make informed trade-off decisions. 
For future work, we plan to incorporate SM ID information and extend the analysis scope to the Grid level to monitor memory access patterns related to hardware bank locations. Additionally, we aim to explore improved data placement strategies and hardware designs for specialized workloads. We will open source \tool{}.
%depending on the paper acceptance .

% This paper presents \tool{}, a novel profiling tool that investigates inefficient memory access patterns in GPU-accelerated applications and provides valuable insights to bridge the gap between user understanding and GPU hardware details.  
% This paper present \tool{}, a novel profiler investigates inefficient memory access patterns in GPU-accelerated applications and provides rich insight to bridge the gap of user understanding and hardware details of GPU. \tool{} can work on fully-optimized binaries without source code, driver, operating system and hardware modification, provides straightforward GUI to guide user understand memory behaviors of complex kernel. With \tool{}, user can optimize their kernel and iterate with guidance, and make trade-off decisions.
% For future work, we add SM id information and extend the scope to Grid scope, to monitor memory access patterns related in hardware bank locations and investigate better data placement methods or hardware design to deal with specialized workloads.

% \begin{acks}
% Memory heat map (MHM), a visualization scheme of the memory behavior, is widely adopted in profiles, performance evaluation, error checking, and bug detection tools. While profiling tools such as Intel VTune Profiler and HeapTrack are supported on CPUs, GPU architectures currently only have limited support for memory profiling through static compile-time tools.
% \end{acks}

\bibliographystyle{IEEEtran}
\bibliography{references}

% \appendix
% \section{Research Methods}
% \subsection{Part One}
% \lipsum[2-4]

\end{document}